\documentstyle[epsf,aps,psfig]{revtex}
\renewcommand{\thefootnote}{\fnsymbol{footnote}}
\begin{document}
\setcounter{footnote}{1}
\begin{center}
{\Large\bf Color superconductivity in weak coupling}
\\[1cm]
Robert D.\ Pisarski $^1$ and Dirk H.\ Rischke $^2$
\\ ~~ \\
{\it $^1$Department of Physics} \\
{\it Brookhaven National Laboratory, Upton, New York 11973, U.S.A.} \\
{\it email: pisarski@bnl.gov}
\\ ~~ \\
{\it $^2$RIKEN-BNL Research Center} \\
{\it Brookhaven National Laboratory, Upton, New York 11973, U.S.A.} \\
{\it email: rischke@bnl.gov}
\\ ~~ \\ ~~ \\
\end{center}
\begin{abstract} 
We derive perturbatively the gap equations for a color-superconducting
condensate with total spin $J=0$ in dense QCD.
At zero temperature, we confirm the results of Son for the dependence of
the condensate on the coupling constant, and compute the prefactor
to leading logarithmic accuracy.  At nonzero temperature, we find that 
to leading order in weak coupling, the
temperature dependence of the condensate is identical to that in 
BCS-like theories.  The condensates for total spin $J=1$ are classified;
to leading logarithmic accuracy these condensates are
of the same order as those of spin $J=0$.
\end{abstract}
\renewcommand{\thefootnote}{\arabic{footnote}}
\setcounter{footnote}{0}

\section{Introduction}

Cooper's theorem \cite{BCS,fetter,Scalapino} implies that if there is an 
attractive interaction in a cold Fermi sea, the system is unstable with 
respect to formation of a particle-particle condensate. 
In QCD, single-gluon exchange between quarks of different
color generates an attractive interaction in the color-antitriplet 
channel \cite{bailinlove}. Thus, it appears unavoidable that color 
superconductivity occurs in dense quark matter which is sufficiently cold
\cite{general,colorflavorlock,prlett,Son,prscalar,hong,continuity,hongetal,SchaferWilczek,parity,prlett2,JMEproc,rockefeller,hsuschwetz,schaferlast,ehhs}.
How a dense quark phase matches onto hadronic matter
is difficult to address \cite{prlett,continuity}.  
In particular, while a quark-quark condensate
may form, such condensation competes
with the tendency of a quark-quark pair to bind with a third quark,
to form a color-singlet hadron.  

One way of understanding color superconductivity is to compute
at very high densities, where by asymptotic freedom, 
perturbation theory can be used. 
At nonzero temperature, but zero quark density, it is well known
that perturbation theory is a particularly bad approximation \cite{temp}.
If $g$ is the coupling constant for QCD, the free energy is
not an expansion in $g^2$, but only in $g$, with a series which
is well behaved only for $g \leq 1$.  In contrast, at
zero temperature and nonzero quark density, the free energy
is an expansion in $g^2 \ln(1/g)$, and appears to be 
well behaved for much larger values of the coupling constant,
up to values of $g \leq 4$ \cite{freenergy}. 
Similar conclusions can be reached by comparing the
gluon ``mass'', $m_g \sim g \mu$ or $\sim gT$ to the chemical potential,
$\mu$, or the temperature, $T$ \cite{prlett2}. Thus for cold, dense
quark matter, perturbation theory might give us information
which not even lattice QCD calculations can provide.

Color superconductivity is rather different from
ordinary superconductivity, as in the model of 
Bardeen, Cooper, and Schrieffer (BCS) \cite{BCS,fetter,Scalapino}.
In BCS-like theories, superconductivity is determined by infrared
divergences which arise in the scattering between two fermions close
to the Fermi surface: the initial fermions, with momenta 
${\bf k}$ and $-{\bf k}$, scatter into a pair
with momenta ${\bf k}'$ and $-{\bf k}'$.  
Summing up bubble diagrams generates an instability which is
only cured by a fermion-fermion condensate.  If the fermions
interact through a point-like four-fermion coupling, though, there is
no correlation between the initial and outgoing momenta, ${\bf k}$
and ${\bf k}'$.  In the gap equation, this implies that the
gap function is constant with respect to momentum, as long
as the momenta are near the Fermi surface.

In QCD, however, scattering through single-gluon exchange strongly
correlates the direction of the in- and out-going quarks:
there is a logarithmic divergence for forward-angle scattering,
$\sim \int d\theta/\theta$.  This extra logarithm from 
forward-scattering implies that the gap is not an exponential in 
$1/g^2$, as in BCS-like theories, but only in $1/g$ 
\cite{prlett,Son,SchaferWilczek,prlett2}.
As a consequence, the gap function is no longer constant as
a function of momentum, even about the Fermi surface. 

The logarithmic divergence for forward-angle scattering
arises because in cold, dense quark matter, 
static, magnetic interactions are not screened through a ``magnetic
mass''. This is very different from a system of hot quarks and gluons.
Over large distances, a hot system is essentially three-dimensional;
gluons in three dimensions have power-like infrared 
divergences, which screen static, magnetic fluctuations
through a magnetic mass $\sim g^2 T$. 
In contrast, loop corrections in cold, dense quark
matter are essentially four-dimensional; infrared divergences
are at worst logarithmic, so that any magnetic mass is
at best $\sim \mu \, \exp(-1/g^2)$, which is much
smaller than the scale for color superconductivity, $\sim \mu\,
\exp(-1/g)$  \cite{prlett,Son}.

The dependence of the zero-temperature
color-superconducting spin-zero condensate
$\phi_0$ on the QCD coupling constant $g$ was first
computed by Son \cite{Son}, who used a beautiful renormalization-group 
analysis to show that 
\begin{equation} \label{Eqphi0}
\phi_0  = 2 \, \frac{b_0}{g^5} \, \mu \,
\exp \left( - \frac{\pi}{2 \,\bar{g}} \right) \,\, .
\end{equation}
Here $\bar{g} = g/(3 \sqrt{2} \pi)$ arises naturally from
the solution, assuming three colors.  
In Son's result, $2\,b_0$ is a pure number of order one.  

In this paper we derive the gap equations for a condensate with
total spin $J=0$ and $N_f=2$ massless quark flavors, 
at an arbitrary temperature $T$. 
Our derivation is perturbative, {\it i.e.}, $g \ll 1$.
In this case there are three scales in cold, dense QCD,
the chemical potential $\mu$, the gluon mass $m_g \sim g \mu$, and
the color-superconducting condensate $\phi_0 \sim \mu\, \exp(-1/g)$, and they
are naturally ordered, $\mu \gg m_g \gg \phi_0$.

We solve the gap equations to ``leading logarithmic accuracy'', by
which we mean the following. In the gap equations, the leading terms
are $\sim \ln (\mu/\phi_0)$. These terms generate the exponential
in $1/g$ in Eq.\ (\ref{Eqphi0}), therefore $\ln(\mu/\phi_0)$ is of order
$1/g$. There are also leading logarithmic terms $\sim \ln(\mu/m_g)$, 
which are $\sim \ln (1/g)$ plus a constant.
The $\ln(1/g)$ gives rise to the prefactor $1/g^5$, while the constant
contributes to $b_0$. For $N_f$ flavors of massless quarks we find
\begin{equation} \label{b0}
b_0 = 256\, \pi^4 \left( \frac{2}{N_f} \right)^{5/2}\, b_0'\,\, .
\end{equation}
There are other terms in the gap equations, which do not arise
from $\ln (\mu/m_g)$. These terms are of order one, and thus of the same order 
as the constant term originating from the logarithm $\ln (\mu/m_g)$.
Hence they contribute in the same way to $b_0$. 
We do not compute these, so that in (\ref{b0})
there is an undetermined constant $b_0'$.

Our results were described previously \cite{prlett2}, 
and agree completely with an independent analysis by 
Sch\"afer and Wilczek \cite{SchaferWilczek};
they also overlap with those of Hong {\it et al.} \cite{hongetal}.
The factor $N_f^{-5/2}$
in Eq.\ (\ref{b0}) originates from the $N_f$ dependence of $m_g$; however,
$b_0'$ will also depend on $N_f$ \cite{schaferlast}.

Even though we do not determine $b_0'$, it is very
interesting that the numerical value of $b_0/b_0'$ is large.
This implies that, for chemical potentials of order $\sim 1$ GeV, 
the gap can be of order $100$ MeV \cite{SchaferWilczek,prlett2}. Such large
values of the gap are in accord with previous estimates obtained within
Nambu--Jona-Lasinio models \cite{general}, and are much larger than 
original estimates by Bailin and Love, $\phi_0 \sim 1$ MeV \cite{bailinlove}.

We then solve the gap equations at nonzero temperature. 
We find a surprising result:
while the detailed form of the gap function is very different
in QCD versus BCS-like theories, the temperature dependence
of the condensate --- the dimensionless ratio
of the condensate at a temperature $T$ to that at zero temperature,
$\phi(T)/\phi_0$, is {\it identical\/} to BCS-type theories.  In particular,
the ratio of the critical temperature, $T_c/\phi_0 \simeq 0.567$, is as
in BCS \cite{BCS,fetter}.  
Our result for $\phi(T)/\phi_0$ is valid to leading order 
in weak coupling, even though we cannot compute 
the overall magnitude of $\phi_0$, {\it i.e.}, $b_0'$.
In our mean-field approximation the transition is of
second order, but the transition can be driven first
order by critical fluctuations 
near the would-be critical point \cite{prlett}.

We then classify the condensates with total spin $J=1$.  
Following our classification of condensates
with $J=0$, which employed projectors for chirality, helicity, and energy,
we show that there are two types of spin-one condensates,
longitudinal and transverse. As an example, we solve the gap equations
for $N_f=1$ and find
\begin{equation} \label{phi1}
\phi_1 = 2\, \frac{b_1}{g^5}\, \mu \, 
\exp\left(-\frac{\pi}{2\, \bar{g}} \right) \;\;\;, \;\;\;
b_1 \equiv \frac{b_0}{b_0'} \, b_1'\,\,. 
\end{equation}
That is, not only is
the parametric dependence of the spin-one condensates on $g$ the same as
for spin zero, as argued originally by Son \cite{Son}, but
to leading logarithmic accuracy, even the constant in front is
the same. We do not expect that $b_1'$, the undetermined constant 
analogous to $b_0'$, is the same. Surely $b_1'$ is smaller than
$b_0'$, and depends on whether the condensate is longitudinal or transverse.
This is very different from BCS-like theories, where condensates
with higher spin are typically exponentially suppressed.  In QCD,
higher-spin condensates are only suppressed by a pure number.  

Nevertheless, one dramatic implication of our results is that 
superconductivity in QCD may be very unlike one's intuition from
non-relativistic systems.  Instead of higher-spin gaps being much
smaller, they may be relatively large, which is important for
phenomenology. Consider, for instance, a quark star
with $u$, $d$, and $s$ quarks. In the limit of high densities, 
when the strange quark mass $m_s$ is negligible, 
the number of $u$, $d$, and $s$ quarks are
equal. Then charge neutrality is automatic, and the preferred
color-superconducting condensates are of spin $J=0$, 
with color-flavor locking \cite{colorflavorlock}.
For realistic densities, however, the chemical potentials for
$u$, $d$, and $s$ will not be equal. The strange quark chemical
potential differs from the up and down quark chemical
potentials due to $m_s \gg m_u,\, m_d$.
The up and down quark chemical potentials differ on account of
charge neutrality. These effects
suppress the color-antitriplet $J=0$ condensates, because they are
composed of quarks with different flavor.
On the other hand, the color-antitriplet $J=1$ condensates may form 
between quarks of the same flavor, 
and are not suppressed if the chemical potentials of the
various flavors differ.

This paper is organized as follows.
In Section \ref{II}, we derive the gap equations for a
spin $J=0$ condensate of $N_f = 2$ massless flavors
at an arbitrary temperature $T$. In Section \ref{III} we solve these
equations to leading logarithmic accuracy, first at $T=0$, and then
at nonzero $T$. In Section \ref{J1} we classify the possible spin
$J=1$ condensates, and solve the gap equations for $N_f =1$.
Section \ref{V} concludes this work with a discussion of higher-order effects
which contribute to $b_0'$ and $b_1'$.
Our units are $\hbar=c=1$, the metric tensor is $g^{\mu \nu} = 
{\rm diag}\,(+,-,-,-)$. 4-vectors are denoted by capital letters,
$K \equiv K^\mu = (k^0,{\bf k})$, and $k \equiv |{\bf k}|$.

\section{The Gap Equations} \label{II}

In general, a color-superconducting 
condensate $\Phi^+_{ij,fg}$ is a $N_c \times N_c$ matrix in
fundamental color space ($i,j = 1, \ldots, N_c$), a $N_f \times N_f$ matrix in
flavor space ($f,g = 1,\ldots, N_f$), and a $4 \times 4$ matrix
in Dirac space \cite{prlett}. As shown in \cite{prlett}, 
for $N_f=2$ a color-antitriplet condensate is a flavor singlet,
\begin{equation} \label{cond}
\Phi^+_{ij,fg} \equiv \epsilon_{fg}\, \Phi^+_{ij} \equiv
\epsilon_{fg}\, \epsilon_{ijk} \, \Phi^+_k \,\, .
\end{equation}
By a global color rotation, we can always choose $\Phi^+_k$ to point in
3--direction in color space, $\Phi^+_k \equiv \delta_{k3}\, \Phi^+$.
For $N_f=3$, even if we assume that the dominant condensate is
a color antitriplet, there is always a small admixture of
a color sextet \cite{JMEproc}. 
The color-antitriplet condensate is a flavor antitriplet, 
and the color sextet a flavor sextet \cite{prlett,schaferlast}.
In this paper, for simplicity we consider only the case $N_f=2$.

The gap equation for a color-superconducting condensate 
of massless fermions was derived in \cite{prscalar} 
[cf.\ Eq.\ (A35) of \cite{prscalar}]. At a nonzero temperature $T$,
this equation reads (we suppress fundamental color
and flavor indices for the moment):
\begin{equation} \label{gapeq}
\Phi^+(K) = g^2 \frac{T}{V} \sum_Q
\bar{\Gamma}_a^\mu \, \Delta^{ab}_{\mu \nu}(K-Q)\,  G_0^{-}(Q) \, 
\Phi^+(Q) \, G^{+}(Q)\, \Gamma_b^\nu\,\, .
\end{equation}
Here, $T/V \sum_Q \equiv T\sum_n \int d^3{\bf q}/(2\pi)^3$ 
in the infinite-volume limit
($n$ labels the Matsubara frequencies $\omega_n \equiv (2n+1) \pi T
\equiv iq_0$), and a summation over Lorentz indices $\mu,\nu$ as
well as adjoint color indices $a,b=1,\ldots, N_c^2-1$ is 
implied; $N_c=3$ is the number of colors. 
$\Delta^{ab}_{\mu \nu}$ is the gluon propagator,
\begin{equation} \label{freeprop}
G_0^\pm(Q) \equiv \left( \gamma \cdot Q \pm \mu \gamma_0 \right)^{-1}
\end{equation}
is the propagator for {\em free}, massless particles (upper sign) or
charge-conjugate particles (lower sign),
\begin{equation} \label{quasiprop}
G^{\pm} \equiv \left\{ \left[ G_0^{\pm} \right]^{-1} - \Sigma^\pm
\right\}^{-1}
\end{equation}
is the propagator for {\em quasiparticles\/} (upper sign) or
charge-conjugate quasiparticles (lower sign), and
\begin{equation} \label{selfenergy}
\Sigma^\pm \equiv \Phi^\mp \, G_0^{\mp}\, \Phi^\pm 
\end{equation}
their self energy arising from the interaction with the
condensate. The charge-conjugate condensate is
\begin{equation} \label{chargeconjcond}
\Phi^- \equiv \gamma_0 \left( \Phi^+ \right)^\dagger \gamma_0 \,\, ,
\end{equation}
and the vertices are
\begin{equation}
\Gamma_a^\mu \equiv T_a \gamma^\mu\,\,\,\, , \,\,\,\,
\bar{\Gamma}_a^\mu \equiv C \, \left( \Gamma_a^\mu \right)^T \, C^{-1} 
\equiv - \gamma^\mu T_a^T \,\, ,
\end{equation}
where $\gamma^\mu$ are the Dirac matrices and $T_a$ the Gell-Mann
matrices, $C = - C^\dagger = -C^T = -C^{-1}= i \gamma^2 \gamma_0$
is the charge conjugation matrix.

In the following, we analyze the flavor, color, and Dirac structure
of the gap equation (\ref{gapeq}). For $N_f = 2$, 
the color-flavor structure of the condensate 
(\ref{cond}) does not mix color and flavor indices, and thus the analysis
of flavor and color can be done separately. This is different for
$N_f=3$, where color rotations are locked to flavor rotations.

\subsection{Flavor structure}

We first discuss the flavor structure of Eq.\ (\ref{gapeq}).
Fundamental color indices will be suppressed for the moment.
The free propagator (\ref{freeprop}) is diagonal in flavor,
\begin{equation}
G_{0 \, fg}^\pm(Q) \equiv \delta_{fg} \, G_0^\pm(Q) \equiv \delta_{fg} \,
\left(\gamma \cdot Q \pm \mu\gamma_0 \right)^{-1}\,\, .
\end{equation}
So is the self energy,
\begin{equation}
\Sigma^+_{fg} = \gamma_0 \left( \Phi^+ \right)^\dagger_{fh}
\gamma_0 \, G_{0\, hi}^- \, \Phi^+_{ig}
=  \epsilon_{hf} \, \delta_{hi} \, \epsilon_{ig}\, 
\gamma_0 \left( \Phi^+ \right)^\dagger \gamma_0 \, G_0^- \, \Phi^+  = 
\delta_{fg}\, \gamma_0 \left( \Phi^+ \right)^\dagger \gamma_0 \, 
G_0^- \, \Phi^+ \equiv \delta_{fg} \, \Sigma^+ \,\, ,
\end{equation}
and thus $G^+_{fg} =  \delta_{fg} \, G^+$. Therefore, both the
left- and right-hand sides of (\ref{gapeq}) are simply proportional to
$\epsilon_{fg}$: the flavor structure of the gap equation is trivial
in QCD with $N_f=2$ flavors, and will thus not be explicitly denoted 
in the following.

\subsection{Color structure}

The free propagator (\ref{freeprop}) is diagonal in the color indices
for the fundamental representation,
\begin{equation}
G_{0 \, ij}^\pm(Q) \equiv \delta_{ij} \, G_0^\pm(Q) \equiv \delta_{ij} \,
\left(\gamma \cdot Q \pm \mu\gamma_0 \right)^{-1}\,\, .
\end{equation}
The self energy $\Sigma^+_{ij}$ is also diagonal, but not all diagonal
elements are equal:
\begin{equation}
\Sigma^+_{ij} = \gamma_0 \left( \Phi^+ \right)^\dagger_{ik}  \gamma_0\,
G_{0\, kl}^- \, \Phi^+_{lj} = \epsilon_{ki3}\,\delta_{kl}\, \epsilon_{lj3}
\, \gamma_0 \left(\Phi^+\right)^\dagger \gamma_0 \,G_0^- \, \Phi^+ 
\equiv \left(\delta_{ij} - \delta_{i3}\, \delta_{j3} \right)\Sigma^+
\end{equation}
--- the self energy for quarks with color 3 vanishes~!
This is easy to understand. Let us first note that, according
to Eq.\ (A17) of \cite{prscalar}, the condensate $\Phi^+$ is
actually proportional to $\langle \psi_C \bar{\psi} \rangle$, 
while $\Phi^- \sim \langle \psi \bar{\psi}_C \rangle$.
($\psi_C \equiv C \bar{\psi}^T$ is the charge-conjugate
fermion field.) Thus, according to Eq.\ (\ref{selfenergy}), the self energy
$\Sigma^+$ arises from the following process: a quark with, let's say, color 1 
annihilates with a corresponding antiquark in $\Phi^+$, 
creating a charge-conjugate quark with color 2. This
quark is propagated with the charge-conjugate propagator $G_0^-$, 
and annihilates with a charge-conjugate antiquark of color 2 in 
$\Phi^-$, whereby a quark with color 1
is emitted. As only quarks with colors 1 and 2 condense, it is not
possible to annihilate and create quarks with color 3 in this process; 
thus the latter do not attain a self energy.

One can now compute the color structure of the quasiparticle propagator,
\begin{equation} \label{quasiprop2}
G^+_{ij} = \left\{ \delta_{ij} \left[G_0^+ \right]^{-1}
- \left(\delta_{ij} - \delta_{i3}\, \delta_{j3} \right)\, \Sigma^+ 
\right\}^{-1} = \left(\delta_{ij} - \delta_{i3}\, \delta_{j3} \right)\,
G^+ + \delta_{i3}\, \delta_{j3}\, G_0^+\,\, ,
\end{equation}
where $G^+ \equiv \left\{ \left[G_0^+ \right]^{-1} - \Sigma^+ \right\}^{-1}$.

When inserted into the gap equation (\ref{gapeq}), 
the terms $\sim \delta_{i3} \, \delta_{j3}$ in (\ref{quasiprop2}) vanish, as 
$\Phi^+_{ij} \sim \epsilon_{ij3}$. The gap equation becomes
\begin{equation}
\epsilon_{ij3}\, \Phi^+(K) = - \left(T^a_{1i}\, T^b_{2j} - T^a_{2i}\, T^b_{1j}
\right) \, g^2 \frac{T}{V} \sum_Q \gamma^\mu \, \Delta^{ab}_{\mu \nu}(K-Q)\, 
G_0^{-}(Q) \, \Phi^+(Q) \, G^{+}(Q)\, \gamma^\nu\,\, .
\end{equation}  From 
the explicit form of the Gell-Mann matrices we now infer that
only gluons with adjoint colors 1, 2, 3, and 8 participate in the
gap equation. 

A color-superconducting condensate $\Phi^+_k
\sim \delta_{k3}$ breaks $SU(3)_c$ to $SU(2)_c$;
gluons 1, 2, and 3 then correspond to the generators
of the unbroken subgroup, and thus remain massless, while
the 8th gluon attains a mass through the Anderson--Higgs effect.
Denoting $\Delta^{11}_{\mu \nu} = \Delta^{22}_{\mu \nu} = 
\Delta^{33}_{\mu \nu} \equiv \Delta_{\mu \nu}$ and 
$\Delta^{88}_{\mu \nu} \equiv \tilde{\Delta}_{\mu \nu}$, 
we obtain
\begin{equation}
\Phi^+(K) = \frac{3}{4} \, 
g^2 \frac{T}{V} \sum_Q \gamma^\mu \, \left[ \Delta_{\mu \nu}(K-Q)
- \frac{1}{9}\, \tilde{\Delta}_{\mu \nu} (K-Q) \right] \, 
G_0^{-}(Q) \, \Phi^+(Q) \, G^{+}(Q)\, \gamma^\nu\,\, ,
\end{equation}
where a common factor $\epsilon_{ij3}$ has been dropped from
both sides of the gap equation.

In a complete treatment of the gap equations the effect of the 
condensate on the gluon propagator has to be included \cite{ehhs}.
In contrast, we use the gluon propagator in the 
``hard dense loop'' (HDL) approximation 
\cite{RDPphysica,LeBellac,finitedens,damping}.
The HDL propagator introduces a gluon mass $m_g \sim g\mu$. 
Since in perturbation theory the scale of the condensate $\phi_0$
is much smaller than $m_g$, it is reasonable to expect that, to 
first approximation, we can neglect the effects of the condensate 
on the gluon propagator. A more detailed explanation is given below.
We consequently assume $\tilde{\Delta}_{\mu \nu} \equiv \Delta_{\mu \nu}$,
and obtain
\begin{equation} \label{gapeq2}
\Phi^+(K) = \frac{2}{3} \, 
g^2 \frac{T}{V} \sum_Q \gamma^\mu \, \Delta_{\mu \nu}(K-Q)\, 
G_0^{-}(Q) \, \Phi^+(Q) \, G^{+}(Q)\, \gamma^\nu\,\, .
\end{equation}

\subsection{Dirac structure}

In \cite{prscalar} we have shown that the gap matrix $\Phi^+$ for
a condensate with total spin $J=0$ has the form
\begin{equation} \label{generalJ0}
J=0: \;\;\;\; 
\Phi^+(K) = \sum_{h=r,\ell}\; \sum_{s=\pm}\; \sum_{e=\pm}
\phi_{hs}^e(K)\,{\cal P}_{hs}^e({\bf k}) \,\, .
\end{equation}
The $4 \times 4$ matrices ${\cal P}_{hs}^e ({\bf k})$ are
defined as
\begin{equation} \label{quasiprojectors}
{\cal P}_{hs}^e ({\bf k}) \equiv {\cal P}_h \, {\cal P}_s({\bf k}) \,
\Lambda^e({\bf k})\,\, ,
\end{equation}
where ${\cal P}_h$ are projectors for chirality, $h=r,\ell$,
\begin{equation} \label{chiralproj}
{\cal P}_r = \frac{1+ \gamma_5}{2}\;\;, \;\;\; 
{\cal P}_\ell = \frac{1-\gamma_5}{2}\,\,, 
\end{equation}
while
\begin{equation}
{\cal P}_s ({\bf k})= \frac{1 +s\, \gamma_5\, \gamma_0 \, \bbox{\gamma} \cdot
\hat{\bf k}}{2}\;\;, \;\;\; s = \pm\;\;,
\end{equation}
are projectors for helicity, and
\begin{equation}
\Lambda^e ({\bf k}) = \frac{1 + e\, (\beta_{\bf k} \, \gamma_0\,
\bbox{\gamma} \cdot \hat{\bf k} + \alpha_{\bf k}\, \gamma_0)}{2}
\;\;, \;\;\; e = \pm \;\;,
\end{equation}
are projectors for energy,
$\beta_{\bf k} \equiv k/E_{\bf k},\, \alpha_{\bf k} \equiv
m/E_{\bf k}, \, E_{\bf k} \equiv \sqrt{k^2 + m^2}$.
While the ${\cal P}_{hs}^e({\bf k})$ are composed of projectors, they
are not projectors themselves [cf.\ Eq.\ (B29) in \cite{prscalar}],
and thus were termed ``quasiprojectors'' in \cite{prscalar}.

In the ultrarelativistic limit, $m=0$, the energy projectors
simplify to
\begin{equation}
m=0: \;\;\;\; \Lambda^e ({\bf k}) = \frac{1 + e\, \gamma_0\,
\bbox{\gamma} \cdot \hat{\bf k}}{2}\,\,.
\end{equation}
This has three major consequences. First, the quasiprojectors
(\ref{quasiprojectors}) become true projectors [cf.\ Eq.\ (B29) of
\cite{prscalar}]. Second, ${\cal P}_{r+}^- =
{\cal P}_{r-}^+ = {\cal P}_{\ell+}^+ = {\cal P}_{\ell-}^- \equiv 0$, 
expressing the fact that right-handed, positive-helicity particles
cannot have negative energy, {\em etc.} [cf.\ Eq.\ (B30) of \cite{prscalar}]. 
Third, {\em either\/} the chirality, {\em or\/} the helicity, {\em or\/} 
the energy projector in (\ref{quasiprojectors}) becomes superfluous
[cf.\ Eq. (B31) of \cite{prscalar}]. In the following, we
omit the helicity projector and use just chirality and energy projectors,
\begin{equation} \label{urlimit}
m=0: \;\;\;\; {\cal P}_{hs}^e ({\bf k}) \longrightarrow {\cal P}_h^e({\bf k})
\equiv {\cal P}_h \, \Lambda^e({\bf k})\,\, .
\end{equation}
Then, (\ref{generalJ0}) simplifies to [cf.\ Eq.\ (8) of \cite{prscalar}]
\begin{equation} \label{gapmatrix}
J=0, \, m=0: \;\;\;\; 
\Phi^+(K) = \sum_{h=r,\ell} \; \sum_{e=\pm} 
\phi_h^e(K)\, {\cal P}_h^e({\bf k}) \,\, .
\end{equation}
The quasiparticle propagator assumes the form 
[cf.\ Eq.\ (15) of \cite{prscalar}]
\begin{equation}
G^+(Q) =  \sum_{h=r,\ell} \; \sum_{e=\pm}
\frac{{\cal P}_h^e({\bf q})}{q_0^2 - 
\left[\epsilon^e_q(\phi_h^e)\right]^2}  
\,  \left[G_0^-(Q) \right]^{-1} \,\, ,
\end{equation}
where $\epsilon^e_q(\phi) \equiv \sqrt{(q - e \mu)^2 + |\phi|^2}$. From 
Eq.\ (26) of \cite{prscalar} we then infer that the
gap equation (\ref{gapeq2}) can be written in the form
\begin{equation} \label{19}
\Phi^+(K) = \frac{2}{3}\, g^2 \frac{T}{V} \sum_Q
\gamma^\mu\, \Delta_{\mu \nu}(K-Q) \, 
\sum_{h=r,\ell} \; \sum_{e=\pm} 
\frac{\phi_h^e(Q)}{q_0^2 - \left[\epsilon^e_q(\phi_h^e)\right]^2}  
\, {\cal P}_{-h}^{-e}({\bf q}) \, \gamma^\nu\,\, ,
\end{equation}
where $-h=\ell$, if $h=r$, and $-h = r$, if $h=\ell$.

With the help of the projectors ${\cal P}_h^e({\bf k})$ one can derive
gap equations for the individual gap functions $\phi_h^e$,
\begin{eqnarray} \label{projectedgapeq}
\phi_h^e(K) & = & \frac{2}{3}\, g^2 \frac{T}{V} \sum_Q
\Delta_{\mu \nu}(K-Q) \, \left\{
\frac{\phi_h^e(Q)}{q_0^2 - \left[\epsilon^e_q(\phi_h^e)\right]^2}  
    \, {\rm Tr} \left[{\cal P}_h^e({\bf k})\, \gamma^\mu \,
{\cal P}_{-h}^{-e}({\bf q})\, \gamma^\nu \right] \right. \nonumber \\
&   & \hspace*{3.6cm} \left.
+\, \frac{\phi_h^{-e}(Q)}{q_0^2- \left[\epsilon^{-e}_q(\phi_h^{-e})\right]^2} 
   \, {\rm Tr} \left[{\cal P}_h^e({\bf k}) \, \gamma^\mu \,
{\cal P}_{-h}^e({\bf q}) \, \gamma^\nu \right] \right\} \,\, .
\end{eqnarray}
To obtain this result we have used ${\cal P}_h \, \gamma^\mu
= \gamma^\mu \, {\cal P}_{-h}$ and ${\cal P}_h \, {\cal P}_{-h} = 0$,
so that the gap equations for right- and left-handed condensates decouple.
This is a consequence of the $U(1)_A$ symmetry of the QCD
Lagrangian, which is expected to be effectively restored at
asymptotically high densities \cite{prlett,parity}.

In Coulomb gauge, the HDL propagator is \cite{RDPphysica,LeBellac}
\begin{equation} \label{Coulombgauge}
\Delta_{00}(P) = \Delta_l(P) + \xi_C \,\frac{p_0^2}{p^4} \,\,\,\, ,
\,\,\,\, 
\Delta_{0i}(P) = \xi_C \, \frac{p_0\, p_i}{p^4} \,\,\,\, ,
\,\,\,\,
\Delta_{ij}(P) = \left(\delta_{ij} - \hat{p}_i \hat{p}_j \right)\,  
\Delta_t(P) + \xi_C\, \frac{\hat{p}_i \, \hat{p}_j}{p^2}\,\,.
\end{equation}
We set the gauge parameter $\xi_C \equiv 0$. We shall show later
that our results are manifestly gauge-invariant to leading logarithmic
accuracy. The longitudinal and transverse propagators $\Delta_{l,t}$ are
defined in Eqs.\ (\ref{longtransprops}) below.

The traces in (\ref{projectedgapeq}) are readily evaluated. We need the terms
\begin{mathletters} \label{traces}
\begin{eqnarray}
{\rm Tr} \left[ {\cal P}_h\, \Lambda^e({\bf k}) \, \gamma_0\,
\Lambda^{\mp e}({\bf q}) \, \gamma_0 \right] & = & 
\frac{1 \pm  \hat{\bf k} \cdot \hat{\bf q}}{2} \,\, , \\
\sum_i {\rm Tr} \left[ {\cal P}_h\, \Lambda^e({\bf k}) \, \gamma_i\,
\Lambda^{\mp e}({\bf q})\, \gamma_i \right] & = & - \, 
\frac{3 \mp \hat{\bf k} \cdot \hat{\bf q}}{2} \,\, , \\
{\rm Tr} \left[ {\cal P}_h\, \Lambda^e({\bf k}) \, \bbox{\gamma}
\cdot \hat{\bf p} \, \Lambda^{\mp e}({\bf q})\, \bbox{\gamma} \cdot \hat{\bf p}
\right] & = & - \, \frac{1\pm \hat{\bf k} \cdot \hat{\bf q}}{2}\, 
\frac{(k\mp q)^2}{p^2} \,\, , 
\end{eqnarray}
\end{mathletters}
where ${\bf p} = {\bf k} - {\bf q}$.
Obviously, the final result is independent of the chirality projector.
We therefore conclude that the gap equations (\ref{projectedgapeq}) for
right- and left-handed gap functions are {\em identical}. This means that
right- and left-handed gaps are equal up to a complex phase factor
$\exp(i\theta)$. Condensation fixes the value of $\theta$ 
and breaks $U(1)_A$ spontaneously. As discussed in \cite{prlett,parity} this
leads to spontaneous breaking of parity.

The gap equations for either right- or left-handed gap functions read
\begin{eqnarray}\label{projectedgapeq2}
\phi_h^e(K) & = & \frac{2}{3}\, g^2 \frac{T}{V} \sum_Q \left\{
\frac{\phi_h^e(Q)}{q_0^2 - \left[\epsilon^e_q(\phi_h^e)\right]^2}  
\, \left[ \Delta_l(K-Q) \, \frac{1 + \hat{\bf k} \cdot \hat{\bf q}}{2}
\right. \right. \nonumber \\
&   &  \left. \hspace*{4.1cm}
+ \,\Delta_t(K-Q) \left( - \frac{3-\hat{\bf k} \cdot \hat{\bf q}}{2}
+ \frac{1 + \hat{\bf k} \cdot \hat{\bf q}}{2} \, 
\frac{(k-q)^2}{({\bf k} - {\bf q})^2} \right) \right]  \nonumber \\
&   & \hspace*{1.6cm} 
+\, \frac{\phi_h^{-e}(Q)}{q_0^2- 
\left[\epsilon^{-e}_q(\phi_h^{-e})\right]^2} 
\, \left[ \Delta_l(K-Q) \, \frac{1 - \hat{\bf k} \cdot \hat{\bf q}}{2}
\right. \nonumber \\
&   &  \left. \left. \hspace*{4.5cm}
+ \,\Delta_t(K-Q) \left( - \frac{3+\hat{\bf k} \cdot \hat{\bf q}}{2}
+ \frac{1 - \hat{\bf k} \cdot \hat{\bf q}}{2} \, 
\frac{(k+q)^2}{({\bf k} - {\bf q})^2} \right) \right] \right\} \,\, .
\end{eqnarray}
The gap equations involve singularities from both the quark and gluon
propagators. The poles of 
$1/\left\{q_0^2 - \left[\epsilon^e_q(\phi_h^e)\right]^2\right\}$
give a residue $\sim 1/ \epsilon^e_q(\phi_h^e)$.
Remember, though, that the quasiparticle energy $\epsilon_q^+$ is very small
near the Fermi surface, $\epsilon_\mu^+ = |\phi_h^+|$. The quasi-antiparticle
energy $\epsilon_q^-$, however, is always larger than $\mu$.
Therefore, in weak coupling, the dominant terms arise from the
quasiparticle poles $q_0 = \pm \epsilon_q^+(\phi_h^+)$ \cite{prscalar},
and the contribution from quasi-antiparticle poles can be neglected.
Consequently, we do not need to solve Eq.\ (\ref{projectedgapeq2})
for the quasi-antiparticle gaps, $\phi_h^-$,
in order to determine the solution for the quasiparticle gaps, $\phi_h^+$. 
In the following,  we drop the subscript $h$ and superscript $+$
to simplify the notation, and denote $\phi_h^+(K) \equiv \phi(K)$.

\subsection{Spectral representations}

To perform the Matsubara sum over quark energies $q_0 = -i (2n+1)\pi T$,
we introduce spectral representations. For the gluon propagators 
\cite{RDPphysica,LeBellac},
\begin{mathletters} \label{longtransprops}
\begin{eqnarray}
\Delta_l(P) & \equiv & - \frac{1}{p^2} + \int_0^{1/T} d \tau \, e^{p_0 \tau}\,
\Delta_l(\tau,{\bf p})  \,\,\,\,\,\,  ,  \,\,\,\,\,\,
\Delta_t(P)  \equiv  \int_0^{1/T} d \tau \, e^{p_0 \tau}\,
\Delta_t(\tau,{\bf p}) \,\, , \\
\Delta_{l,t}(\tau,{\bf p}) & \equiv & \int_0^\infty d\omega \, 
\rho_{l,t} (\omega,{\bf p}) 
\left\{ \left[1+n_B(\omega/T)\right] \, e^{-\omega \tau}
+ n_B(\omega/T) \, e^{\omega \tau} \right\} \,\, , 
\end{eqnarray}
\end{mathletters}
where $n_B(x) \equiv 1/(e^x-1)$ is the Bose--Einstein distribution
function. 
The term $-1/p^2$ in the longitudinal propagator
cancels the contribution of $\Delta_l(P)$ at $p_0 \rightarrow \infty$
\cite{LeBellac}.
The spectral densities are given by \cite{RDPphysica,LeBellac}
\begin{mathletters} \label{specdens}
\begin{eqnarray}
\rho_{l,t}(\omega,{\bf p}) & = & \rho^{\rm pole}_{l,t} 
(\omega,{\bf p})\, \delta \left[\omega - \omega_{l,t}({\bf p})\right] + 
\rho^{\rm cut}_{l,t}(\omega,{\bf p}) \, \theta(p-\omega) \,\, , \\
\rho_l^{\rm pole}(\omega,{\bf p}) & = & \frac{\omega\, (\omega^2 - p^2)}{
p^2\,(p^2+3m_g^2-\omega^2)} \,\, , \\
\rho_l^{\rm cut}(\omega,{\bf p}) & = & \frac{2M^2}{\pi}\, \frac{\omega}{p}\,
\left\{ \left[ p^2+3\, m_g^2 \, \left( 1 -\frac{\omega}{2p} \, 
\ln \left| \frac{p+\omega}{p-\omega} \right| \right) \right]^2 + 
\left(2 M^2 \,\frac{\omega}{p} \right)^2 \right\}^{-1} \,\, ,\\
\rho_t^{\rm pole}(\omega,{\bf p}) & = & \frac{\omega\, (\omega^2 - p^2)}{
3m_g^2\,\omega^2-(\omega^2-p^2)^2} \,\, , \\
\rho_t^{\rm cut}(\omega,{\bf p}) & = & \frac{M^2}{\pi}\, \frac{\omega}{p}\,
\frac{p^2}{p^2-\omega^2}\,
\left\{ \left[ p^2+\frac{3}{2}\, m_g^2 \, 
\left( \frac{\omega^2}{p^2-\omega^2} +
\frac{\omega}{2p}\, \ln \left| \frac{p+\omega}{p-\omega} \right|
\right) \right]^2 +
\left(M^2\, \frac{\omega}{p} \right)^2 \right\}^{-1} \, \, .
\end{eqnarray}
\end{mathletters} 
The basic parameter of the HDL propagators is the gluon mass
\begin{equation} \label{gluonmass}
m_g^2 \equiv  N_f \, \frac{g^2 \mu^2}{6 \pi^2}  +  
\left( N_c + \frac{N_f}{2} \right) \frac{g^2 T^2}{9} \,\, .
\end{equation}
We also found it convenient to introduce
\begin{equation}
M^2 \equiv \frac{3 \pi}{4} \, m_g^2 \simeq 2.36 \, m_g^2\,\, .
\end{equation}
The functions $\omega_{l,t}({\bf p})$ are the solutions of the equations 
\begin{mathletters} \label{dispersion}
\begin{eqnarray}
p^2 + 3m_g^2 \left[ 1 -\frac{\omega_l}{2p} \, \ln \left(
\frac{\omega_l + p}{\omega_l -p} \right) \right] 
& = & 0 \,\, , \\
p^2 \left( \omega_t^2 -p^2\right) - 
\frac{3}{2}\,m_g^2\, \left[ \omega_t^2 + \frac{\omega_t}{2p} 
\left(p^2-\omega_t^2\right)\,
\ln \left( \frac{\omega_t+p}{\omega_t-p} \right) \right] & = & 0\,\, ,
\end{eqnarray}
\end{mathletters}
and define the dispersion relations
for longitudinal and transverse gluons, respectively.
They satisfy $\omega_{\l,t} ({\bf p}) \geq m_g$.

We argue later that the phase-space region which dominates the
gap integrals is the nearly static, small-momentum limit.
The gluon energies are on the order of the gap, $\omega \sim \phi$,
while the gluon momenta $p$ are much larger than $\omega$,
but much smaller than $m_g$. In this limit, $\omega \ll p \ll m_g$,
\begin{equation} \label{approxspec}
\rho_l^{\rm cut} (\omega,{\bf p}) \simeq \frac{2M^2}{\pi}\,
\frac{\omega}{p}\, \frac{1}{(p^2+ 3m_g^2)^2}\;\;\;\;,\;\;\;\;
\rho_t^{\rm cut} (\omega,{\bf p}) \simeq \frac{M^2}{\pi}\,
\frac{\omega \, p}{ p^6 +\left(M^2 \omega \right)^2} \,\, .
\end{equation}

We also introduce a spectral representation for the quantity
\begin{equation}
\Xi(Q) \equiv \frac{\phi(Q)}{q_0^2- 
\left[\epsilon_q(\phi)\right]^2}
\equiv \int_0^{1/T} d\tau \, e^{q_0 \tau}\, \Xi(\tau,{\bf q}) \,\,.
\end{equation}
Neglecting the singularities of $\phi(Q)$ in the complex $q_0$ plane,
and assuming $\phi(q_0,{\bf q})$ to be an even function of $q_0$, we find
\begin{equation}
\Xi(\tau,{\bf q}) \equiv \int_0^\infty d \omega\, \tilde{\rho}
(\omega,{\bf q}) \left\{ \left[1-n_F(\omega/T)\right]\, e^{-\omega \tau} -
 n_F(\omega/T) \, e^{\omega \tau} \right\} \,\, ,
\end{equation}
where
\begin{equation} \label{fermionspec}
\tilde{\rho}(\omega,{\bf q}) \equiv - \frac{ 
\phi(\omega,{\bf q})}{
2 \, \omega} \, \delta(\omega - \epsilon_q) \,\, ,
\;\;\;\; \epsilon_q \equiv \epsilon_q(\phi)\,\, ,
\end{equation}
and $n_F(x) = 1/(e^x +1)$ is the Fermi--Dirac distribution function.

By neglecting the singularities of $\phi(Q)$, the only contribution
to the spectral representation of $\Xi(Q)$ is from the
poles of $1/\left\{q_0^2 - \left[\epsilon_q(\phi)\right]^2\right\}$, which
generates the delta function in the spectral density (\ref{fermionspec}).
This forces the energy in the gap function $\phi(q_0,{\bf q})$ to lie on the
quasiparticle mass shell, $q_0 = \epsilon_q(\phi)$.

The Matsubara sums over $q_0$ can now be computed as (${\bf p} \equiv
{\bf k} - {\bf q})$
\begin{mathletters} \label{matsu}
\begin{eqnarray}
\lefteqn{T \sum_{q_0} \Delta_l(K-Q) \, \Xi(Q)  = 
- \frac{\phi(\epsilon_q,{\bf q})}{2\, \epsilon_q} \left\{
- \frac{1}{2}\, \tanh \left( \frac{\epsilon_q}{2T} \right) \, 
\frac{2}{p^2} +
\int_0^\infty d \omega\, \rho_l(\omega,{\bf p}) \right.} \nonumber \\
& \times & \left[ \;\; 
\frac{1}{2} \, \tanh \left( \frac{\epsilon_q}{2T}\right) \,
\left( \frac{1}{k_0+\omega+\epsilon_q} 
- \frac{1}{k_0-\omega-\epsilon_q}
- \frac{1}{k_0-\omega+\epsilon_q} 
+ \frac{1}{k_0+\omega-\epsilon_q} \right)
\right. \nonumber \\
&   & \left. + \left.  \frac{1}{2} \, \coth \left( \frac{\omega}{2T}\right) \,
\left( \frac{1}{k_0+\omega+\epsilon_q} 
- \frac{1}{k_0-\omega-\epsilon_q}
+ \frac{1}{k_0-\omega+\epsilon_q} 
- \frac{1}{k_0+\omega-\epsilon_q} \right)
\right] \right\} \,\, , \label{matsu1} \\
\lefteqn{T \sum_{q_0} \Delta_t(K-Q) \, \Xi(Q)  = 
- \frac{\phi(\epsilon_q,{\bf q})}{2\, \epsilon_q} 
\int_0^\infty d \omega\, \rho_t(\omega,{\bf p})} \nonumber\\
& \times  & \left[ \;\;
\frac{1}{2} \, \tanh \left( \frac{\epsilon_q}{2T}\right) \,
\left( \frac{1}{k_0+\omega+\epsilon_q} 
- \frac{1}{k_0-\omega-\epsilon_q}
- \frac{1}{k_0-\omega+\epsilon_q} 
+ \frac{1}{k_0+\omega-\epsilon_q} \right)
\right. \nonumber \\
&   & + \left. \frac{1}{2} \, \coth \left( \frac{\omega}{2T}\right) \,
\left( \frac{1}{k_0+\omega+\epsilon_q} 
- \frac{1}{k_0-\omega-\epsilon_q}
+ \frac{1}{k_0-\omega+\epsilon_q} 
- \frac{1}{k_0+\omega-\epsilon_q} \right) \right]\,\, , \label{matsu2} 
\end{eqnarray}
\end{mathletters}
where we have made use of (\ref{fermionspec}) and 
$e^{k_0/T} \equiv -1$, since $k_0 = -i (2n+1) \pi T$.

\subsection{Analytic continuation}

In field theories, the only physical quantities are those on the
mass shell, such as S-matrix elements.
In our case, this implies that we need the gap function
not for Euclidean values of the energy, $k_0 = -i (2n+1) \pi T$,
but on the quasiparticle mass shell, 
for real values of $k_0 = \epsilon_k$. This is achieved
by analytically continuing $k_0 \rightarrow \epsilon_k + i \eta$
\cite{fetter}.
The analytic continuation of the gap equation (\ref{projectedgapeq2}) 
introduces a term $i \eta$ in the energy denominators 
in (\ref{matsu}). As $1/(x+i \eta) \equiv {\rm P} (1/x) -i \pi \delta(x)$ 
(where ${\rm P}$ stands for the principal value), this generates 
an imaginary part for the function $\phi (\epsilon_k,{\bf k})$.
As shown below, this imaginary part is down by $g$ relative to the
real part. This justifies neglecting the singularities of
$\phi(Q)$ in deriving (\ref{matsu}).
In the following, implicitly we take the principal value of
all energy denominators.
Finally, since the gap function $\phi(k_0,{\bf k})$ 
does not depend on the orientation of the 3-vector ${\bf k}$, but only on
its magnitude, $k$, cf.\ (\ref{projectedgapeq2}), the on-shell
gap function is a function of $k$ only,
$\phi(\epsilon_k,{\bf k}) \equiv \phi(\epsilon_k, k)\equiv \phi_k$.

\subsection{Contribution from longitudinal gluons}

The contribution from longitudinal gluons is given by Eq.\ (\ref{matsu1}). 
In general, this equation can only be evaluated numerically, using the 
full expressions (\ref{specdens}) for the spectral densities. 
We can proceed analytically by noting that, due to the 
factor $1/\epsilon_q$, the integral over momenta ${\bf q}$ is dominated
by the region close to the Fermi surface. In the following,
we shall therefore assume that the momenta ${\bf k}$ and ${\bf q}$ are close
to the Fermi surface, such that $\epsilon_k \sim \epsilon_q  \ll m_g$.
Now rewrite
\begin{eqnarray}
\lefteqn{ \frac{1}{\omega+\epsilon_q+\epsilon_k} 
+ \frac{1}{\omega+\epsilon_q-\epsilon_k}
+ \frac{1}{\omega-(\epsilon_q+\epsilon_k)} 
+ \frac{1}{\omega-(\epsilon_q-\epsilon_k)} }\nonumber \\
& = &
\frac{1}{\omega}
\, \left[ 4 
- \frac{\epsilon_q+\epsilon_k}{\omega+\epsilon_q+\epsilon_k}
- \frac{\epsilon_q-\epsilon_k}{\omega+\epsilon_q-\epsilon_k}
+\frac{\epsilon_q+\epsilon_k}{\omega-(\epsilon_q+\epsilon_k)} 
+ \frac{\epsilon_q-\epsilon_k}{\omega-(\epsilon_q-\epsilon_k)} 
 \right] \,\, ,
\label{relat}
\end{eqnarray}
and make use of the sum rule \cite{RDPphysica,LeBellac}
\begin{equation}
2 \int_0^\infty d \omega \, \frac{\rho_l(\omega,{\bf p})}{\omega}
= \frac{1}{p^2} + \Delta_l(0,{\bf p})\;\;,\;\;\;\;
\Delta_l(0,{\bf p}) \equiv - \frac{1}{p^2 +3m_g^2}\,\,,
\end{equation}
to obtain
\begin{equation} \label{long}
T \sum_{q_0} \Delta_l(K-Q) \, \Xi(Q)  = 
- \frac{\phi(\epsilon_q,{\bf q})}{2\, \epsilon_q} \left\{
 \frac{1}{2}\, \tanh \left( \frac{\epsilon_q}{2T} \right) \, \left[
-\frac{2}{p^2+3m_g^2} +
{\cal J}_l({\bf k},{\bf q}) \right] + 
{\cal K}_l({\bf k},{\bf q}) \right\} \,\,,
\end{equation}
where
\begin{equation} \label{Jl}
{\cal J}_l({\bf k},{\bf q}) \equiv
\int_0^\infty d \omega\, \frac{\rho_l(\omega,{\bf p})}{\omega}
\, \left[ \frac{\epsilon_q+\epsilon_k}{\omega-(\epsilon_q+\epsilon_k)} 
- \frac{\epsilon_q+\epsilon_k}{\omega+\epsilon_q+\epsilon_k}
+ \frac{\epsilon_q-\epsilon_k}{\omega-(\epsilon_q-\epsilon_k)} 
- \frac{\epsilon_q-\epsilon_k}{\omega+\epsilon_q-\epsilon_k} \right]\,\, ,
\end{equation}
and
\begin{equation} \label{Kl}
{\cal K}_l({\bf k},{\bf q}) \equiv
\int_0^\infty d \omega\, \rho_l(\omega,{\bf p})
\, \frac{1}{2} \, \coth \left( \frac{\omega}{2T}\right) \,
\left[ \frac{1}{\omega+\epsilon_q+\epsilon_k} 
+ \frac{1}{\omega+\epsilon_q-\epsilon_k}
- \frac{1}{\omega-(\epsilon_q+\epsilon_k)} 
- \frac{1}{\omega-(\epsilon_q-\epsilon_k)} \right]
\,\, .
\end{equation}
The integral ${\cal J}_l$ consists of two parts, one from
the pole term in the spectral density, and one from the
cut term. The pole term is
\begin{equation}
{\cal J}_l^{\rm pole}({\bf k},{\bf q}) =
\frac{1}{p^2}\, \frac{\omega_l^2 -p^2}{p^2+3m_g^2 -\omega_l^2}\, 
\left[ \frac{\epsilon_q+\epsilon_k}{\omega_l-(\epsilon_q+\epsilon_k)} 
- \frac{\epsilon_q+\epsilon_k}{\omega_l+\epsilon_q+\epsilon_k}
+ \frac{\epsilon_q-\epsilon_k}{\omega_l-(\epsilon_q-\epsilon_k)} 
- \frac{\epsilon_q-\epsilon_k}{\omega_l+\epsilon_q-\epsilon_k} \right]\,\, .
\end{equation}
As $\omega_l \geq m_g \gg \epsilon_k \sim \epsilon_q$, we may
expand the energy denominators around $\omega_l$,
\begin{equation} \label{Jlpole}
{\cal J}_l^{\rm pole}({\bf k},{\bf q}) \simeq
\frac{1}{p^2}\, \frac{\omega_l^2 -p^2}{p^2+3m_g^2 -\omega_l^2}\, 
4\,\frac{\epsilon_q^2+\epsilon_k^2}{\omega_l^2}\,\,.
\end{equation}
This contribution is quadratic in $\epsilon_q/\omega_l \sim
\epsilon_k/\omega_l \ll 1$ and thus negligible. 

The cut term is estimated using the approximate form of the spectral
density (\ref{approxspec}),
\begin{eqnarray}
{\cal J}_l^{\rm cut}({\bf k}, {\bf q}) & \simeq &
\frac{2 M^2}{\pi p}\, \frac{1}{\left(p^2 + 3m_g^2\right)^2}
\int_0^{p} d \omega\, 
\left[ \frac{\epsilon_q+\epsilon_k}{\omega-(\epsilon_q+\epsilon_k)} 
- \frac{\epsilon_q+\epsilon_k}{\omega+\epsilon_q+\epsilon_k}
+ \frac{\epsilon_q-\epsilon_k}{\omega-(\epsilon_q-\epsilon_k)} 
- \frac{\epsilon_q-\epsilon_k}{\omega+\epsilon_q-\epsilon_k} \right] 
\nonumber \\
& = & \frac{2M^2}{\pi p}\, \frac{1}{\left(p^2 + 3m_g^2\right)^2}
\left[ (\epsilon_q+\epsilon_k) \, \ln \left|\frac{\epsilon_q+\epsilon_k-p}{
\epsilon_q+\epsilon_k+p}\right| + (\epsilon_q-\epsilon_k)\,
\ln \left| \frac{\epsilon_q-\epsilon_k-p}{\epsilon_q-\epsilon_k+p}\right|
\right] \,\, .
\end{eqnarray}
This integral is proportional to factors $\epsilon_q \pm
\epsilon_k$, which parametrically cancel
the prefactor $1/\epsilon_q$ in (\ref{long}). 
(This is true except for $p\rightarrow 0$, where 
these factors cancel when expanding the logarithms.
However, the Jacobian of the angular integration in the 
gap equation, $\int d \cos\theta \sim \int dp\, p$,
suppresses this contribution.) 
It is thus negligible compared to the term 
$\sim 2/(p^2 + 3m_g^2)$ in (\ref{long}).

We now turn to evaluate ${\cal K}_l({\bf k},{\bf q})$.
After expanding the energy denominators in (\ref{Kl}) around
$\omega_l \geq m_g \gg \epsilon_k \sim \epsilon_q$, the pole part reads
\begin{equation}\label{Klpole}
{\cal K}_l^{\rm pole}({\bf k},{\bf q}) \simeq
- \frac{1}{p^2}\, \frac{\omega_l^2 -p^2}{p^2+3m_g^2-\omega_l^2}\,
\frac{1}{2}\, \coth \left( \frac{\omega_l}{2T} \right) \, 4\, 
\frac{\epsilon_q}{\omega_l} \,\, .
\end{equation}
The range of temperatures of interest are limited by $T_c$, the
critical temperature for the onset of color superconductivity.
As $T_c \sim \phi_0 \ll m_g \leq \omega_l$, $\coth(\omega_l/2T) \sim 1$.
The factor $\epsilon_q$ cancels the prefactor in (\ref{long}), such
that ${\cal K}_l^{\rm pole}$ is negligible compared to the 
term $\sim 2/(p^2 + 3m_g^2)$ in (\ref{long}).

To estimate the cut term, we use the approximate expression
(\ref{approxspec}), and employ a relation similar to (\ref{relat}),
\begin{eqnarray}
\lefteqn{ \frac{1}{\omega+\epsilon_q+\epsilon_k} 
+ \frac{1}{\omega+\epsilon_q-\epsilon_k}
- \frac{1}{\omega-(\epsilon_q+\epsilon_k)} 
- \frac{1}{\omega-(\epsilon_q-\epsilon_k)} }\nonumber \\
& = &
- \frac{1}{\omega} \, \left[
\frac{\epsilon_q+\epsilon_k}{\omega+\epsilon_q+\epsilon_k}
+\frac{\epsilon_q-\epsilon_k}{\omega+\epsilon_q-\epsilon_k}
+\frac{\epsilon_q+\epsilon_k}{\omega-(\epsilon_q+\epsilon_k)} 
+\frac{\epsilon_q-\epsilon_k}{\omega-(\epsilon_q-\epsilon_k)} 
 \right]\,\, , \label{relat2}
\end{eqnarray}
to obtain
\begin{eqnarray}
{\cal K}_l^{\rm cut}({\bf k},{\bf q}) & \simeq &
- \frac{2M^2}{\pi p}\, \frac{1}{(p^2+3m_g^2)^2}
\int_0^p d \omega\, \frac{1}{2} \, \coth \left(\frac{\omega}{2T}
\right) \nonumber \\
& \times & \left[
\frac{\epsilon_q+\epsilon_k}{\omega+\epsilon_q+\epsilon_k}
+\frac{\epsilon_q-\epsilon_k}{\omega+\epsilon_q-\epsilon_k}
+\frac{\epsilon_q+\epsilon_k}{\omega-(\epsilon_q+\epsilon_k)} 
+\frac{\epsilon_q-\epsilon_k}{\omega-(\epsilon_q-\epsilon_k)} 
 \right]\,\,  .
\end{eqnarray}
For $\omega$ of order $T$ or larger, this contribution is small, since
then $\coth(\omega/2T) \sim 1$, and thus all terms
are proportional to factors $\epsilon_q \pm \epsilon_k$.
For $\omega \ll T$, however, the large (classical) occupation
number density of gluons can enhance this contribution.
In the classical limit, one approximates $\coth (\omega/2T) \simeq 2T/\omega$.
Simultaneously, as the range of validity of this approximation
is $\omega \ll T$, the upper limit of the integral should now be replaced
by $p^* \equiv {\rm min}(T,p)$.
Reverting the step (\ref{relat2}), we obtain
\begin{equation}
{\cal K}_l^{\rm cut}({\bf k},{\bf q}) \simeq
 \frac{2 M^2\,T}{\pi p}\,  \frac{1}{(p^2+3m_g^2)^2} 
\, \ln \left| \frac{p^*+\epsilon_q+\epsilon_k}{p^*-(\epsilon_q+\epsilon_k)}
\, \frac{p^*+\epsilon_q-\epsilon_k}{p^*-(\epsilon_q-\epsilon_k)}
\right|\,\, .
\end{equation}
For $T\rightarrow 0$, we may expand the logarithm to show that
this contribution is quadratically small in $T$.
On the other hand, for $T \rightarrow T_c$, close to the Fermi surface
$\epsilon_q \sim \epsilon_k \sim \phi \rightarrow 0$, and the logarithm
can be expanded for $p^* \gg \epsilon_q \pm \epsilon_k$. The result
is proportional to $\epsilon_q \pm \epsilon_k$, which cancels the
prefactor $1/\epsilon_q$, and thus suppresses this contribution.
We shall therefore neglect it in the following.

The final result for the longitudinal contribution is thus
\begin{equation} \label{longfinal}
T \sum_{q_0} \Delta_l(K-Q) \, \Xi(Q)  \simeq 
\frac{\phi(\epsilon_q,{\bf q})}{2\, \epsilon_q}\,
 \frac{1}{2}\, \tanh \left( \frac{\epsilon_q}{2T} \right) \,
\frac{2}{p^2+3m_g^2}\,\, .
\end{equation}
This result could also have been obtained by simply taking
the static limit of the longitudinal gluon propagator,
$\Delta_l(0,{\bf p})$, on the left-hand side in (\ref{longfinal}), 
and performing the Matsubara sum directly. 
We conclude that the contribution of static electric gluons
dominates over that of non-static electric gluons.
We note that, while the individual terms (\ref{Jlpole}) and (\ref{Klpole})
exhibit an apparent infrared divergent prefactor $1/p^2$, the sum
of all terms is infrared finite, as can be shown by computing
the contribution of electric gluons without using (\ref{relat}).

\subsection{Contribution from transverse gluons}

The contribution from transverse gluons is written in a form
similar to (\ref{long}):
\begin{equation} \label{trans}
T \sum_{q_0} \Delta_t(K-Q) \, \Xi(Q)  = 
- \frac{\phi(\epsilon_q,{\bf q})}{2\, \epsilon_q} \left[
\frac{1}{2}\, \tanh \left( \frac{\epsilon_q}{2T} \right) \,
{\cal J}_t({\bf k},{\bf q})  + 
{\cal K}_t({\bf k},{\bf q}) \right] \,\,,
\end{equation}
where
\begin{equation} \label{Jt}
{\cal J}_t({\bf k},{\bf q}) \equiv
\int_0^\infty d \omega\, \rho_t(\omega,{\bf p})
\, \left[ \frac{1}{\omega+\epsilon_q+\epsilon_k}
+\frac{1}{\omega+\epsilon_q-\epsilon_k}
+\frac{1}{\omega-(\epsilon_q+\epsilon_k)}
+\frac{1}{\omega-(\epsilon_q-\epsilon_k)} \right]\,\, ,
\end{equation}
and
\begin{equation} \label{Kt}
{\cal K}_t({\bf k},{\bf q}) \equiv
\int_0^\infty d \omega\, \rho_t(\omega,{\bf p})
\, \frac{1}{2} \, \coth \left( \frac{\omega}{2T}\right) \,
\left[ \frac{1}{\omega+\epsilon_q+\epsilon_k} 
+ \frac{1}{\omega+\epsilon_q-\epsilon_k}
- \frac{1}{\omega-(\epsilon_q+\epsilon_k)} 
- \frac{1}{\omega-(\epsilon_q-\epsilon_k)} \right]
\,\, .
\end{equation}
The function ${\cal J}_t$ consists of a pole and a cut term.
In the first, we expand the energy denominators around
$\omega_t \geq m_g \gg \epsilon_q \sim \epsilon_k$, to obtain
to leading order in $\epsilon_q \pm \epsilon_k$:
\begin{equation} \label{complicated}
{\cal J}_t^{\rm pole}({\bf k},{\bf q}) \simeq
4\, \frac{\omega_t^2 - p^2}{3m_g^2 \omega_t^2 - (\omega_t^2 - p^2)^2}\,\,.
\end{equation}

For the cut term, we again employ (\ref{approxspec}) to obtain
\begin{equation}
{\cal J}_t^{\rm cut}({\bf k},{\bf q}) \simeq
\frac{M^2 p}{\pi} \int_0^p d \omega\, \frac{\omega}{p^6+\left(
M^2 \omega \right)^2}\,
\left[\frac{1}{\omega+\epsilon_q+\epsilon_k} 
+ \frac{1}{\omega+\epsilon_q-\epsilon_k}
+ \frac{1}{\omega-(\epsilon_q+\epsilon_k)} 
+ \frac{1}{\omega-(\epsilon_q-\epsilon_k)} \right]
\,\, .
\end{equation}
The $\omega$ integral can be performed analytically. Denote
$a \equiv p^3/M^2$. Then
\begin{eqnarray}
{\cal J}_t^{\rm cut}({\bf k},{\bf q}) & \simeq & 
\frac{p}{\pi\, M^2} \int_0^p d \omega\, \frac{\omega}{\omega^2 + a^2}
\, \left[\frac{1}{\omega+\epsilon_q + \epsilon_k} 
+ \frac{1}{\omega+\epsilon_q - \epsilon_k}
+ \frac{1}{\omega-(\epsilon_q + \epsilon_k)} 
+ \frac{1}{\omega-(\epsilon_q - \epsilon_k)} \right] \nonumber \\
& = & \frac{p}{\pi\, M^2}  \left\{
2 \left[\frac{a}{a^2+(\epsilon_q + \epsilon_k)^2} 
+\frac{a}{a^2+(\epsilon_q - \epsilon_k)^2}\right]
\arctan \left( \frac{p}{a} \right) \right. \nonumber \\
&   & \left. \hspace*{1.2cm}
- \frac{\epsilon_q + \epsilon_k}{a^2+(\epsilon_q + \epsilon_k)^2}\, 
\ln \left| \frac{p+\epsilon_q + \epsilon_k}{p-(\epsilon_q + \epsilon_k)}\right|
- \frac{\epsilon_q - \epsilon_k}{a^2+(\epsilon_q - \epsilon_k)^2}\, 
\ln \left| \frac{p+\epsilon_q - \epsilon_k}{p-(\epsilon_q - \epsilon_k)}\right|
 \right\} \,\, . \label{52}
\end{eqnarray}
The terms proportional to $\epsilon_q \pm \epsilon_k$ are of
higher order and can be neglected, such that
\begin{equation}
{\cal J}_t^{\rm cut}({\bf k},{\bf q})  \simeq 
\left[\frac{p^4}{p^6 + M^4 (\epsilon_q + \epsilon_k)^2}
+ \frac{p^4}{p^6 + M^4 (\epsilon_q - \epsilon_k)^2} \right]
\frac{2}{\pi}\,\arctan\left(\frac{M^2}{p^2} \right) \,\, .
\end{equation}
The $\arctan$ cuts ${\cal J}_t$ off for momenta $p$ larger than 
$\sim M$. To make further progress,
in the following we replace the more complicated $\arctan$
with a simple $\theta$ function cutoff, $2\,\arctan (M^2/p^2)/\pi
\rightarrow \theta(M - p)$. However, to retain consistency
with the sum rule \cite{RDPphysica,LeBellac}
\begin{equation} \label{sumrule}
2 \int_0^\infty d\omega \, \frac{\rho_t(\omega,{\bf p})}{\omega} 
= \frac{1}{p^2}\,\, ,
\end{equation}
we also have to modify the result for the pole term (\ref{complicated}).
To enact the modification, note that ${\cal J}_t/2$ is identical to
the left-hand side of the sum rule (\ref{sumrule}), when 
we set $\epsilon_q = \epsilon_k = 0$ in (\ref{Jt}).
Thus, the most simple choice is
\begin{equation}
{\cal J}_t^{\rm pole} ({\bf k},{\bf q}) \simeq \frac{2}{p^2}\, 
\theta(p-M)\,\, ,
\end{equation}
since then, for $\epsilon_q = \epsilon_k = 0$,
\begin{equation}
{\cal J}_t({\bf k},{\bf q}) = {\cal J}_t^{\rm pole} ({\bf k},{\bf q})
+{\cal J}_t^{\rm cut}({\bf k},{\bf q}) \equiv \frac{2}{p^2}\,\, .
\end{equation}

The function ${\cal K}_t$ also consists of a pole and a cut term.
In the first, we expand the energy denominators around $\omega_t
\geq m_g \gg \epsilon_q \sim \epsilon_k$,
\begin{equation}
{\cal K}_t^{\rm pole} ({\bf k}, {\bf q}) \simeq
- \frac{\omega_t^2 - p^2}{3m_g^2 \omega_t^2 - (\omega_t^2-p^2)^2}\,
\frac{1}{2}\, \coth \left(\frac{\omega_t}{2T} \right)
\, \frac{4\, \epsilon_q}{\omega_t}
\,\,.
\end{equation}
As the factor $\epsilon_q$ cancels the prefactor $1/\epsilon_q$ in
(\ref{trans}), this contribution is of higher order and thus
negligible.

The cut term is estimated with (\ref{approxspec}) to be
(as before, $a = p^3/ M^2$)
\begin{eqnarray}
{\cal K}_t^{\rm cut} ({\bf k},{\bf q}) & \simeq &
\frac{p}{\pi\, M^2} \int_0^p d \omega \,
\frac{\omega}{\omega^2+a^2}\, \frac{1}{2}\,
\coth \left( \frac{\omega}{2T} \right) \nonumber \\
& \times &  \left[
\frac{1}{\omega + \epsilon_q + \epsilon_k} 
+ \frac{1}{\omega+ \epsilon_q - \epsilon_k}
- \frac{1}{\omega - (\epsilon_q + \epsilon_k)} 
- \frac{1}{\omega - (\epsilon_q - \epsilon_k)} \right] \,\,.
\end{eqnarray}
For $\omega$ of order $T$ or larger, 
$\coth (\omega/2T) \simeq 1$, similar steps
that led to (\ref{52}) now show that the final result is
proportional to $\epsilon_q \pm \epsilon_k$, and thus
of higher order. For $\omega \ll T$, however, 
\begin{eqnarray}
{\cal K}_t^{\rm cut} ({\bf k},{\bf q}) & \simeq &
\frac{p\, T}{\pi \, M^2} \int_0^{p^*} d \omega \,
\frac{1}{\omega^2+a^2}\, \left[
\frac{1}{\omega + \epsilon_q + \epsilon_k} 
+ \frac{1}{\omega+ \epsilon_q - \epsilon_k}
- \frac{1}{\omega - (\epsilon_q + \epsilon_k)} 
- \frac{1}{\omega - (\epsilon_q - \epsilon_k)} \right]
\nonumber \\
& \simeq & \frac{M^2}{\pi} \left[ \frac{p\, T}{p^6 + M^4 
(\epsilon_q + \epsilon_k)^2}
\, \ln \left| \frac{p^* + \epsilon_q + \epsilon_k}{p^*-
(\epsilon_q + \epsilon_k)} \right|
+ \frac{p\, T}{p^6 + M^4 (\epsilon_q - \epsilon_k)^2}
\, \ln \left| \frac{p^* + \epsilon_q - \epsilon_k}{p^*-
(\epsilon_q - \epsilon_k)} \right|\right]\,\, , 
\end{eqnarray}
where as before $p^* = {\rm min}(T,p)$, and
terms proportional to $\epsilon_q \pm \epsilon_k$ 
have been neglected in the last line. 
For small $T$, an expansion of the logarithms shows
that this contribution is quadratic in $T$. On the other hand, for
$T \rightarrow T_c$, an expansion of the logarithms for
$p^* \gg \epsilon_q \pm \epsilon_k$ shows that it is proportional
to $\epsilon_q \pm \epsilon_k$; it thus parametrically cancels
the prefactor $1/\epsilon_q$. We shall thus neglect
this contribution in the following.

Our final result for the transverse contribution is therefore
\begin{eqnarray}\label{transfinal}
\lefteqn{T \sum_{q_0} \Delta_t(K-Q) \, \Xi(Q) } \nonumber \\
& \simeq  & - \frac{\phi(\epsilon_q,{\bf q})}{2\, \epsilon_q}\,
 \frac{1}{2}\, \tanh \left( \frac{\epsilon_q}{2T} \right) \,
\left\{ \frac{2}{p^2}\, \theta(p-M)
+ \theta(M-p) \left[ \frac{p^4}{p^6 + M^4(\epsilon_q+\epsilon_k)^2}
+ \frac{p^4}{p^6+ M^4(\epsilon_q-\epsilon_k)^2} \right] \right\} \,\, .
\end{eqnarray}

\subsection{The gap equation}

Collecting the results from the previous subsections, and replacing
the angular integration $\int d \cos \theta$ by an integral
over $p = |{\bf k} - {\bf q}|$, the gap equation (\ref{projectedgapeq2})
reads
\begin{eqnarray}
\lefteqn{\phi_k  =  \frac{g^2}{24 \pi^2\, k}\, 
\int_{\mu-\delta}^{\mu + \delta} dq \, \frac{q}{\epsilon_q}
\, \tanh \left(\frac{\epsilon_q}{2T} \right) \, \phi_q
\int_{|k-q|}^{k+q} dp \; p \left\{  \frac{2}{p^2 + 3m_g^2}\,
\frac{(k+q)^2 - p^2}{4 k q}  \right.} \nonumber \\
&    & \left. + \left[ \frac{2}{p^2} \,
\theta(p-M) + 
\theta(M-p) \left( \frac{p^4}{p^6+M^4
(\epsilon_q+\epsilon_k)^2} + \frac{p^4}{p^6+M^4
(\epsilon_q-\epsilon_k)^2}\right) \right] \left( 1 + \frac{p^2}{4 k q} -
\frac{(k^2-q^2)^2}{4 k q p^2} \right) \right\}\,\,. \label{particlegap2}
\end{eqnarray}
Here, we restricted the $q$ integration to a region $\mu - \delta
\leq q \leq \mu + \delta,\, \delta \ll \mu$ around the Fermi surface, 
since we assumed that $\epsilon_q \sim \epsilon_k \ll m_g \sim g \mu \ll \mu$.
Therefore, we may set $k \simeq q \sim \mu$. 
Also, to leading order we may neglect $p^2$ terms with respect to $(k+q)^2$
terms. 
Then, the gap equation (\ref{particlegap2}) simplifies to
\begin{equation} \label{gapeqalmostfinal}
\phi_k  \simeq  \frac{g^2}{24 \pi^2}\, 
\int_{\mu-\delta}^{\mu + \delta} \frac{dq}{\epsilon_q}
\, \tanh \left(\frac{\epsilon_q}{2T} \right) \, \phi_q
\, \left[  \ln \left(\frac{4 \mu^2}{3m_g^2} \right)
 + \ln \left( \frac{4\mu^2}{M^2} \right) + 
\frac{1}{3} \ln \left( \frac{M^2}{|\epsilon_q^2-\epsilon_k^2|}
\right) \right] \,\,.
\end{equation}
The first term in brackets, $\sim \ln (4\mu^2/3m_g^2)$, arises from static 
electric gluons [proportional to $2/(p^2 + 3m_g^2)$ in Eq.\
(\ref{particlegap2})], the second, $\sim \ln(4\mu^2/M^2)$, 
from non-static magnetic gluons [proportional to 
$\theta(p-M)$ in Eq.\ (\ref{particlegap2})], and the last term,
$\sim \ln(M^2/|\epsilon_q^2-\epsilon_k^2|)$, from
soft, Landau-damped magnetic gluons [proportional to 
$\theta(M-p)$ in Eq.\ (\ref{particlegap2})]. These terms can be combined
to give
\begin{equation} \label{gapeqfinal}
\phi_k  \simeq  \frac{g^2}{18 \pi^2}\, 
\int^{\delta}_{0} \frac{d(q-\mu)}{\epsilon_q}
\, \tanh \left(\frac{\epsilon_q}{2T} \right) \, \frac{1}{2} \,
\, \ln \left(\frac{b^2\mu^2}{|\epsilon_q^2-\epsilon_k^2|} \right)\, \phi_q
\,\,,
\end{equation}
where we exploited the symmetry of the integrand in (\ref{gapeqalmostfinal})
around the Fermi surface to restrict the $q-\mu$ integration to
the interval $\mu \leq q \leq \mu+\delta$, {\it i.e.}, we only
integrate over momenta $q$ from the Fermi surface up to 
$\mu + \delta$. Moreover, we have defined
\begin{equation} \label{b}
b \equiv 256\, \pi^4 \left( \frac{2}{N_f g^2} \right)^{5/2}\,\, .
\end{equation}
The temperature dependence of $m_g$, cf.\ Eq.\ (\ref{gluonmass}),
has been neglected, as $T \sim \phi_0 \ll m_g \sim g \mu$.

The result (\ref{b}) for $b$ agrees with the analysis of 
Sch\"afer and Wilczek \cite{SchaferWilczek}.
It does not agree with that of Hong {\it et al.} \cite{hongetal},
because they neglected the contribution of electric gluons.
Note that our gap equation (\ref{gapeqfinal}) differs from that of
Son \cite{Son} and of Sch\"afer and Wilczek \cite{SchaferWilczek}
in that we integrate over momenta $q$, while they integrate over
Euclidean energies $q_0$. The energy dependence of our gap function 
is always given by the quasiparticle mass shell $\phi(K) = \phi(\epsilon_k,
{\bf k})$. One advantage of our approach is that the extension to
nonzero temperature is immediate.

{}From Eq.\ (\ref{particlegap2}), the gluon momenta which dominate
the contribution of nearly static, transverse gluons, are
$p^6 \sim M^4 (\epsilon_q \pm \epsilon_k)^2$, or for
energies $\epsilon_q,\, \epsilon_k$ close to the Fermi surface,
$p \sim (m_g^2 \phi)^{1/3}$. While these momenta are small relative
to the gluon mass $m_g$, they are much larger than the 
condensate $\phi$. This is why we can use an HDL
propagator in the gap equations, neglecting the effects of the condensate.

The approximations which lead to (\ref{gapeqfinal}) are only valid to
leading logarithmic accuracy, in that we neglect terms of order
1 relative to $\ln(\mu/m_g) \sim \ln(1/g)$.
Possible contributions of order 1, which contribute to $b_0'$ in
Eq.\ (\ref{b0}), are discussed in Section \ref{V}.

\section{Solving the gap equation} \label{III}

The gap equation (\ref{gapeqfinal}) is an integral equation for the
function $\phi_k \equiv \phi(\epsilon_k,{\bf k})$. In general, it
can only be solved numerically \cite{SchaferWilczek}. 
In this section, we first discuss the parametric dependence
of the solution to (\ref{gapeqfinal}) on the QCD coupling constant.
To this end, it is instructive to understand which terms
in (\ref{gapeqalmostfinal}) determine the exponent 
and the prefactor in Eq.\ (\ref{Eqphi0}). We then
derive an approximate analytical solution by converting the
integral equation (\ref{gapeqfinal})
into a differential equation. The solution is
discussed in detail at zero and at nonzero temperature.

\subsection{Parametric dependence of the solution on the QCD coupling constant}
\label{IIIA}

Let us introduce the variable
\begin{equation}
\bar{g} \equiv \frac{g}{3 \sqrt{2}\, \pi} \,\,.
\end{equation} 
At $T=0$, the general structure of the gap equation (\ref{gapeqalmostfinal}) 
is
\begin{equation} \label{generalstruc}
\phi_k =  \frac{\bar{g}^2}{2}\; \int_0^\delta  \frac{d(q-\mu)}{\epsilon_q} \;
\left[ \, \ln \left( \frac{\mu^2}{|\epsilon_q^2 - \epsilon_k^2|}\right)
 +  2 \, \ln \left(\frac{b_0}{b_0'g^5} \right) + c \,\right] \; \phi_q
\; . 
\end{equation}
The first term, $\sim \ln(\mu^2/|\epsilon_q^2-\epsilon_k^2|)$, 
arises from nearly static, transverse gluons. As compared to
(\ref{gapeqalmostfinal}) we have factored out a term
$\sim \ln(M^2/\mu^2)$.
The second term, $\sim \ln[b_0/(b_0'g^5)]$, combines this term with the
contribution from electric and non-static magnetic gluons. The ratio
$b_0/b_0'$ is a well-defined number and given in Eq.\ (\ref{b0}). 
These first two terms comprise the leading logarithmic approximation.
The last term, $c$, represents contributions that go beyond
leading logarithmic accuracy, such as terms
of order one in $\int d(q-\mu)/\epsilon_q$, or of higher order,
for instance $\sim \int d(q-\mu)$. We have neglected these terms
in our derivation of Eq.\ (\ref{gapeqalmostfinal}).
In Eq.\ (\ref{b0}), they were written in terms of
the undetermined constant $b_0'$.
We are therefore permitted to choose $c \equiv 2 \ln b_0'$ in the following.

In order to understand which terms are responsible for the exponent
and the prefactor in Eq.\ (\ref{Eqphi0}) let us
solve (\ref{generalstruc}) assuming that the gap function $\phi_q$
is constant as a function of momentum. This will give the wrong
coefficient in the exponent, but suffices to understand the
parametric dependence of (\ref{Eqphi0}) on $g$.
For $k= \mu,\, \phi_q = \phi_k = \phi_0 = const.$, the $(q-\mu)$ integral
in Eq.\ (\ref{generalstruc}) can be performed to yield ($\delta \gg \phi_0$)
\begin{equation} \label{quadraticeq}
1 \simeq \bar{g}^2\, \left[ \frac{1}{2}\, \ln^2  \left( 
\frac{2 \delta}{\phi_0}\right) + 
\ln \left(\frac{b_0 \mu}{g^5 \delta} \right) \, 
\ln  \left( \frac{2 \delta}{\phi_0}\right) \right]\,\, .
\end{equation}
The first term arises from the leading term $\sim \ln (\mu^2/|\epsilon_q^2
-\epsilon_k^2|)$ in (\ref{generalstruc}), while the second comes from
the leading logarithmic term $\sim \ln[b_0/(b_0'g^5)]$.
The quadratic equation (\ref{quadraticeq}) for
$\ln (2\delta/\phi_0)$ has the solution
\begin{equation} \label{quadsol}
\ln \left( \frac{2\delta}{\phi_0} \right)
\simeq -\ln \left( \frac{b_0 \mu}{g^5 \delta} \right)
+ \sqrt{\ln^2 \left( \frac{b_0\mu}{g^5 \delta}\right) + 
\frac{2}{\bar{g}^2}} \,\,.
\end{equation}
In weak coupling, the term $\sim 2/\bar{g}^2$ 
dominates the right-hand side, so that we can expand the square root. 
This term then gives rise to the exponent in $1/\bar{g}$ for
$\phi_0$. The term $\sim \ln [b_0 \mu/(g^5 \delta)]$ in (\ref{quadsol}) 
gives rise to the prefactor of the exponential.
In this way we obtain
\begin{equation} \label{phi0wrong}
\phi_0 \simeq 2 \, \frac{b_0}{g^5} \, \mu \, \exp \left( 
-\frac{\sqrt{2}}{\bar{g}} \right) \times \left[1+O(g) \right] \,\, .
\end{equation}
Note that, to leading order in $g$, 
the dependence on the cutoff $\delta$ cancels in the final
result. Equation (\ref{phi0wrong}) is rather similar to Eq.\ (\ref{Eqphi0}).
The difference is that, due to our (erroneous) assumption of a
constant gap function, the coefficient in the
exponent is incorrect, $\sqrt{2}$, instead of $\pi/2$.

\subsection{Converting the integral equation into a differential equation}

The gap equation (\ref{gapeqfinal}) can be solved analytically
by approximating the logarithm under the integral. 
As in \cite{Son}, we replace
\begin{equation} \label{sonstrick}
\frac{1}{2}\, \ln \left(\frac{b^2\mu^2}{|\epsilon_q^2-\epsilon_k^2|} \right)
\rightarrow  \ln \left( \frac{b \mu}{\epsilon_q} \right) \,
\theta(q-k) + \ln \left( \frac{b \mu}{\epsilon_k} \right) \,
\theta(k-q) \,\, ,
\end{equation}
and the gap equation (\ref{gapeqfinal}) becomes
\begin{equation} 
\phi_k  \simeq  \bar{g}^2\, \left[ 
\ln \left( \frac{b \mu}{\epsilon_k} \right)
\int_{0}^{k-\mu} \frac{d(q-\mu)}{\epsilon_q}
\, \tanh \left(\frac{\epsilon_q}{2T} \right) \, \phi_q
+\int_{k-\mu}^{\delta} \frac{d(q-\mu)}{\epsilon_q}
\, \tanh \left(\frac{\epsilon_q}{2T} \right) \, 
\ln \left( \frac{b \mu}{\epsilon_q} \right)\,\phi_q \right]
\,\,.
\end{equation}
Upon differentiation with respect to $k$, we see that the gap function
$\phi_k$ monotonously decreases from its maximum
at the Fermi surface $k=\mu$
(we assume $\phi_q \geq 0$ for $0 \leq q-\mu \leq \delta$).  
Let us introduce the variables
\begin{mathletters}
\begin{eqnarray} \label{defvary}
y & \equiv & \ln \left( \frac{2\,b\mu}{q-\mu + \epsilon_q} \right) \,\,, \\
x & \equiv & \ln \left( \frac{2\,b\mu}{k-\mu + \epsilon_k} \right) \,\,, 
\label{defvarx} \\
x^* & \equiv & \ln \left( \frac{2\,b\mu}{\epsilon_\mu} 
\right) \equiv \ln \left(\frac{2\, b\mu}{\phi} \right)\,\, ,
\end{eqnarray}
\end{mathletters}
where $\phi \equiv \phi_\mu \equiv \phi(\epsilon_\mu,\mu \hat{\bf k})$
is the value of the gap at the Fermi surface. 
Obviously, $x^* \geq x,\, y$.
As $\delta \ll m_g \ll \mu \ll b \mu$ [cf.\ (\ref{b})], 
both $x$ and $y$ are much larger than 1. 
In terms of these new variables,
\begin{mathletters}
\begin{eqnarray} \label{eq}
\epsilon_q & = & b\mu\, e^{-y} \left[1 + \left(\frac{\phi(y)}{\phi}\, 
e^{-(x^*-y)}\right)^2\right] \equiv \epsilon(y)  \,\,, \\
\epsilon_k & = & b\mu\, e^{-x} \left[1 + \left(\frac{\phi(x)}{\phi}\, 
e^{-(x^*-x)} \right)^2 \right] = \epsilon(x)\,\, , \label{ek}
\end{eqnarray}
\end{mathletters}
where we defined a new function $\epsilon(y)$.

The transformation to the new variable $y$ is natural, because 
\begin{equation}
dy = - \frac{d(q-\mu)}{\epsilon_q}
\end{equation}
is the measure for integration, without any further Jacobian.
It is similar to Son's variable $y_{\rm Son} = \ln (\mu/q_0)$.
For Son's variable, however, $dy_{\rm Son} = -dq_0/q_0$, so that in
the gap equation, it is necessary to include a Jacobian for the 
transformation from $q_0$ to $y_{\rm Son}$. This Jacobian does not
affect his results for the parametric dependence of the gap function on $g$,
but to leading logarithmic accuracy, it does affect the prefactor and 
the shape of the gap function, cf.\ discussion at the end of subsection
\ref{IIIC}.

In the new variables $x,\,y$ the approximate gap equation reads
\begin{equation} \label{approxgapeq}
\phi(x) \simeq  \bar{g}^2 \left[x \int_x^{x^*} dy\; \tanh \left(
\frac{\epsilon(y) }{2T} \right) \, \phi(y) 
+ \int_{\ln (b\mu/\delta)}^x d y \; y \, \tanh \left( 
\frac{\epsilon(y)}{2T} \right) \, \phi(y) \right] \,\, ,
\end{equation}
with $\epsilon(y)$ given by (\ref{eq}).  Here, we have neglected
terms $\sim \ln \left[1+(\phi(y)/\phi)^2 e^{-2(x^*-y)}\right] \ll y$ and
$\sim \ln \left[1+(\phi(x)/\phi)^2 e^{-2(x^*-x)}\right]\ll x$. This
is a good approximation, as the value of the logarithms is bounded from above
by $\ln 2 \simeq 0.693$, while $x, \, y \geq \ln (b\mu/\delta) \gg 1$.
Differentiating with respect to $x$ yields
\begin{mathletters}
\begin{eqnarray} 
\frac{d}{dx}\,\phi(x) & \simeq & \bar{g}^2 \int_x^{x^*} dy \; \tanh \left(
\frac{ \epsilon(y) }{2T} \right) \, \phi(y) \,\, ,\\
\frac{d^2}{dx^2}\, \phi(x) & \simeq & - \bar{g}^2 \,\tanh \left(
\frac{ \epsilon(x) }{2T} \right) \, \phi(x) \,\, .\label{diffeq}
\end{eqnarray}
\end{mathletters}
This is the generalization of Son's equation for the gap function
to nonzero temperature.

\subsection{Zero temperature} \label{IIIC}

The approximation (\ref{sonstrick}) has succeeded to convert the
original integral equation into the differential equation (\ref{diffeq}). 
For nonzero $T$, however, because $\epsilon(x)$ is a complicated
function of $x$, this equation still requires a numerical
solution. On the other hand, for $T=0$ the solution is simply
a trigonometric function. Determining its phase and amplitude from
the values of $\phi(x^*)$ and $d\phi/dx$ at $x^*$, we obtain
\begin{equation} \label{solution}
\phi(x) = \phi_0 \, \cos\left[\bar{g}(x^*-x)\right] \,\, .
\end{equation}
The value of the zero-temperature gap function at the Fermi 
surface, $\phi_0$, can now be obtained by inserting the solution 
(\ref{solution}) into the approximate gap equation for $x = x^*$. 
This yields
\begin{equation}
\phi_0 = 2\, \delta \, \exp \left[ - \frac{1}{\bar{g}}\, \arctan 
\left( \frac{1}{\bar{g} \, \ln ( b\mu/\delta)} \right) \right]\,\, .
\end{equation} 
The dependence on the momentum cutoff $\delta$ is spurious: in
weak coupling, $\bar{g} \ll 1$, we may expand the $\arctan$ with the
result
\begin{equation}
\phi_0 = 2\, b \mu\, \exp \left( -\frac{\pi}{2 \, \bar{g}} \right)
\times \left[ 1 + O(\bar{g}^2) \right] \,\, .
\end{equation}
This is identical to Eq.\ (\ref{Eqphi0}), except that here the undetermined
constant $b_0'$ is equal to one, since we computed only to 
leading logarithmic accuracy and dropped terms of order one in the
gap equation.

\begin{figure} 
\hspace*{2.5cm} 
\psfig{figure=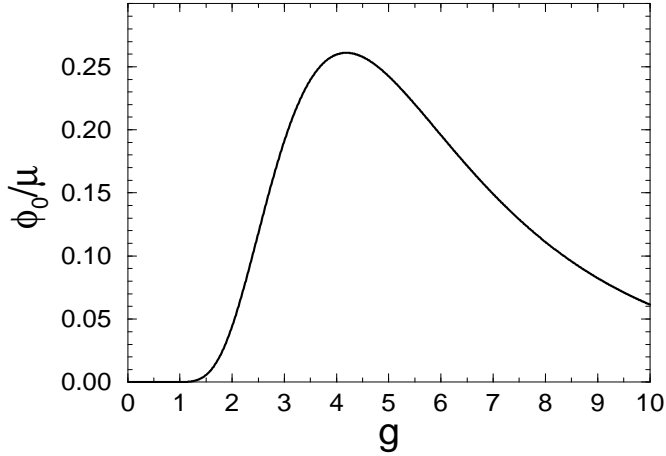,width=2in,height=2.5in,angle=-90}
\vspace*{-1cm}
\caption{$\phi_0/\mu$ as function of $g$. We have set the undetermined
constant $b_0'$ in Eq.\ (\ref{b0}) equal to $1$.}
\label{phi0}
\end{figure}

While our analysis is strictly valid only for small values of $g$,
it is instructive to extrapolate to strong coupling.
The behavior of $\phi_0$ as function of $g$ is shown in Fig.\
\ref{phi0}. One observes that $\phi_0$ has a maximum at a coupling
constant $g \simeq 4.2$. The maximum value is quite large,
$\phi_0 \simeq 0.26\, \mu$, due to the large prefactor
$b_0$, Eq.\ (\ref{b0}), which has important implications for
phenomenology \cite{prlett2}.
\newpage

To leading order in $\bar{g}$, $x^* \equiv \pi/(2 \bar{g})$, such that
\begin{equation}
\phi(x) = 2 \, b \mu \, \exp \left( -\frac{\pi}{2\, \bar{g}} \right)
\, \sin (\bar{g} x) \,\, .
\label{eqresult}
\end{equation}
The gap function peaks at the Fermi surface, $x=x^*$, and varies
over a region $x^* - x \sim 1/\bar{g}$. 
This implies that the gap integral is dominated by momenta
exponentially close to the Fermi surface, $\epsilon_q \sim b\mu
\, e^{-y} \ll m_g$. For example, when $q-\mu \sim m_g$, and 
consequently $y \sim 1$, $\sin(\bar{g}y) \sim \bar{g}y$,
and the gap function is $\bar{g}$ times smaller than at the Fermi surface.
Physically, this dependence of the gap function on momentum reflects
the fact that single-gluon exchange is dominated by
forward-angle scattering \cite{prlett2}.

The effect of using the variable $y$, Eq.\ (\ref{defvary}),
instead of a variable $z \equiv \ln (b\mu/\epsilon_q)$ analogous
to Son's variable $y_{\rm Son} \equiv \ln (\mu/q_0)$ 
\cite{Son,SchaferWilczek} is the appearance
of a prefactor $2$ in Eq.\ (\ref{eqresult}), as well as a factor
of $2$ under the logarithm in $x^* = \ln (2b\mu/\phi)$. 
These factors of 2 were found empirically 
in a numerical analysis by Sch\"afer and Wilczek \cite{SchaferWilczek}.

\subsection{Nonzero temperature}

At nonzero temperature, the gap function depends on both $x$ and $T$.
In this subsection we denote this dependence by $\phi(x,T)$. 
The value of $\phi(x,T)$ at the Fermi surface is denoted by
$\phi(T)$. As before, $\phi_0 = \phi(0)$.

In weak coupling, we can compute $\phi(x,T)$ by
assuming $\phi(x,T) \simeq \phi(T)\, \phi(x,0)/\phi_0$.
In other words, the $x$ dependence of the gap function is taken to
be the same as at zero temperature, so that the only effect
of nonzero temperature is to change the overall magnitude of the gap
function. We first present this calculation, and then 
discuss why, in weak coupling, it is reasonable to assume that the
gap function as function of $x$ does not change when the temperature is 
varied. Finally, we verify this assumption by numerical calculations.

For reference, we review how the solution to a BCS-type
gap equation changes with
temperature.  Taking the BCS coupling constant to be $G$,
the gap equation is of the form
\begin{equation}
\phi(T) = G^2 \int^{\Lambda}_0 \frac{d(q-\mu)}{\epsilon_q} \tanh 
\left(\frac{\epsilon_q}{2 T}\right) \phi(T) \;\;\; , \;\;\;
\epsilon_q = \sqrt{(q-\mu)^2 + \phi(T)^2} \; .
\label{bcsgap}
\end{equation}
In BCS-type theories, the momentum dependence of the gap equation
can be ignored, so that we just have a fixed-point equation which
determines $\phi(T)$. Besides the trivial solution, $\phi(T)=0$,
at sufficiently small temperatures there is also a non-trivial solution
$\phi(T) \neq 0$. It is determined by solving the equation
\begin{equation} \label{bcsgap0}
1 = G^2 \int^{\Lambda}_0 \frac{d(q-\mu)}{\epsilon_q} \tanh 
\left(\frac{\epsilon_q}{2 T}\right) \,\,.
\end{equation}
At zero temperature, the solution is
\begin{equation} \label{solutionbcs}
1 \simeq G^2 \ln \left( \frac{2 \Lambda}{\phi_0} \right) \;\;\;\; ,
\;\;\;\; \phi_0 \simeq 2 \, \Lambda \, \exp \left( - \frac{1}{G^2} \right) \; .
\end{equation}
The gap equation requires us to introduce
a cutoff, $\Lambda$, but all that matters is that the cutoff
is much larger than the value of the gap at zero temperature, 
$\Lambda \gg \phi_0$.  

With increasing $T$, the thermal factor $\tanh (\epsilon_q/2T)$ reduces the
integrand in (\ref{bcsgap0}), so that $\phi(T)$ has to decrease 
in order to achieve equality of right- and left-hand sides of Eq.\ 
(\ref{bcsgap0}).
Above a critical temperature, $T_c$, this is no longer possible and
we only have the trivial solution, $\phi(T) = 0,\, T \geq T_c$.

Let us investigate in more detail how the balance
between the left- and right-hand sides of Eq.\ (\ref{bcsgap0})
is achieved at nonzero temperature. We divide the integration region in 
(\ref{bcsgap0}) into two pieces, for $q-\mu$ smaller or larger 
than $\kappa \phi_0$, where $\kappa$ is some pure number, which
is assumed to be large. The first region corresponds to momenta near 
the Fermi surface, $0 \leq q - \mu \leq \kappa \phi_0$.
Although $\kappa \gg 1$, because the gap is {\em exponentially small\/} in
the coupling constant, Eq.\ (\ref{solutionbcs}), in weak coupling
this region constitutes only a rather small contribution to
the complete gap integral in (\ref{bcsgap0}).

The bulk of the integral comes from the region of momenta 
away from the Fermi surface, $\kappa \phi_0 \leq q - \mu \leq \Lambda$.
The quintessential point of our argument, which we shall apply to the
QCD case shortly, is that,
for temperatures on the order of $\phi_0$, the thermal factor
can be neglected in this region,
up to corrections $\sim \exp(-\kappa \phi_0/T)$. The gap equation
(\ref{bcsgap0}) can therefore be written in the form:
\begin{equation} \label{bcsgap000}
1 \simeq G^2 \left[ \int_0^{\kappa \phi_0} \frac{d(q-\mu)}{\epsilon_q}
\, \tanh \left( \frac{\epsilon_q}{2T} \right)
 + \int_{\kappa \phi_0}^\Lambda \frac{d(q-\mu)}{\epsilon_q} \right] \,\, .
\end{equation}
In the small region very near the Fermi surface, 
for momenta $0 \leq q-\mu \leq \kappa \phi_0$,
the thermal factor must be retained.

In weak coupling, $G \ll 1$, the region away from the Fermi surface 
almost saturates the $1$ on the
left-hand side of the gap equation (\ref{bcsgap000}). 
To see this, we compute the
contribution of this region to the right-hand side,
\begin{equation}
G^2 \int_{\kappa \phi_0}^\Lambda \frac{d(q- \mu)}{\epsilon_q}
\simeq G^2 \, \ln \left( \frac{\Lambda}{\kappa \phi_0} \right)
\simeq 1 - G^2 \, \ln ( 2 \kappa )\,\, ,
\end{equation}
where we used the solution (\ref{solutionbcs}) at $T=0$.
Hence, in order to fulfill the gap equation (\ref{bcsgap000}),
the first integral in (\ref{bcsgap000}) has
to balance the term $G^2 \, \ln (2 \kappa)$,
\begin{equation} \label{balance}
G^2\, \ln (2 \kappa) \simeq G^2 
\int_0^{\kappa \phi_0} \frac{d(q-\mu)}{\epsilon_q}
\, \tanh \left( \frac{\epsilon_q}{2T} \right)\,\, .
\end{equation}
This equation can be written in a more concise form.
Note that
\begin{equation}  \label{2kappa}
\int_0^{\kappa \phi_0} \frac{d(q-\mu)}{\epsilon_q^0} \simeq 
\ln (2 \kappa)\,\, ,
\end{equation}
where $\epsilon_q^0 = \sqrt{(q-\mu)^2 + \phi_0^2}$ is the
quasiparticle excitation energy at zero temperature.
We therefore replace the term $\ln (2 \kappa)$ in (\ref{balance}) by
this integral,
\begin{equation} \label{bcsgap00}
0 \simeq G^2 \int_0^{\kappa \phi_0} d(q-\mu) \left[\frac{1}{\epsilon_q}
\, \tanh \left( \frac{\epsilon_q}{2T}\right) 
- \frac{1}{\epsilon_q^0} \right] \,\,.
\end{equation}
The same result could have been obtained directly, by
subtracting Eq.\ (\ref{bcsgap0}) at zero
temperature from (\ref{bcsgap0}) at $T \neq 0$, and using the fact
that, for momenta $q-\mu \geq \kappa \phi_0$,
$\epsilon_q \simeq \epsilon_q^0$, so that 
the contributions to the integrals away from the Fermi surface
cancel in the subtraction process.
The integral in (\ref{bcsgap00}) is finite in the infrared,
even for $\phi(T) = 0$, since $\tanh(x)/x \rightarrow 1$ as $x \rightarrow 0$.

Dividing all quantities with dimension of energy by $\phi_0$, 
one realizes that Eq.\ (\ref{bcsgap00}) determines the ratio
$\phi(T)/\phi_0$ as a function of $T/\phi_0$. Note that
this ratio is independent of the cutoff $\Lambda$, as well as the 
coupling constant $G$, since $G^2$ is just an overall constant in
(\ref{bcsgap00}) which we can divide out.

Equation (\ref{bcsgap00}) has the following interpretation.
As mentioned above, the $1$ on the left-hand side of the gap equation 
(\ref{bcsgap000}) is almost completely saturated by momenta away from the 
Fermi surface, $\kappa \phi_0 \leq q- \mu \leq \Lambda$,
where thermal corrections are negligible.
Thermal effects become important in a small region near the
Fermi surface, $0 \leq q- \mu \leq \kappa \phi_0$.
Only in this region, the change of the gap function with temperature 
has to compensate for the presence of the thermal factor.
The ratio $\phi(T)/\phi_0$ is therefore {\em completely\/}
determined by investigating how the gap equation changes with
$T$ in a small region around the Fermi surface~!

In Eq.\ (\ref{bcsgap00}), we are allowed to send $\kappa \rightarrow \infty$,
as done in Eq.\ (10) of \cite{prlett2}, because
$\tanh x  \rightarrow 1$ for $x \rightarrow \infty$, and $\epsilon_q
\rightarrow \epsilon_q^0$ for $q-\mu \rightarrow \infty$. 
The ratio $\phi(T)/\phi_0$ is therefore not only independent of the
cutoff $\Lambda$, and the coupling constant $G$, but also of $\kappa$.
However, sending $\kappa \rightarrow \infty$ somewhat obscures the fact 
that only a small region around the Fermi surface controls
how $\phi(T)$ changes with temperature.

Equation (\ref{bcsgap00}) cannot be evaluated analytically for arbitrary $T$,
although it is easy to solve numerically; as is well known, 
the transition is of second order, with $\phi(T) \rightarrow 0$
as $T \rightarrow T_c$. At the critical temperature $T_c$,
however, one can evaluate (\ref{bcsgap00}) analytically:
\begin{equation}
\int^\infty_0 d(q-\mu)
\left[ \frac{1}{\epsilon_q} \; \tanh\left(\frac{\epsilon_q}{2 T_c} \right)
- \frac{1}{\epsilon^0_q } \right]
\simeq \ln \left(\frac{\zeta\, \phi_0}{2\, T_c}\right) = 0
\;\;\; , \;\;\; \zeta = \frac{2 e^\gamma}{\pi} \simeq 1.13 \; .
\label{bcsgapT4}
\end{equation}
Here $\gamma \simeq 0.577$ is the Euler-Mascheroni constant.
The solution is 
\begin{equation}
\frac{T_c}{\phi_0} = \frac{\zeta}{2} = 0.567 \; .
\label{bcsgapT5}
\end{equation}
which is the usual result in BCS-theory \cite{BCS,fetter}. 

We now go through a similar computation for QCD, neglecting
the change in the $x$ dependence of the gap function
with temperature, $\phi(x,T) \simeq \phi(T)\, \phi(x,0)/\phi_0$. 
Consider the gap equation (\ref{approxgapeq}) for $x = x^*$, {\it i.e.},
for the gap function at the Fermi surface, $\phi(T)$.
Assuming $T < T_c$, and taking the non-trivial solution
$\phi(T) \neq 0$, we can divide both sides by $\phi(T)$. We obtain
\begin{equation}
1 \simeq \bar{g}^2 \int_{\ln(b\mu/\delta)}^{x^*} dy \; y\, 
\tanh\left(\frac{\epsilon(y)}{2T}\right) \, \frac{\phi(y,0)}{\phi_0}
\,\, .
\label{qcdgapT1}
\end{equation}
This is the equation analogous to (\ref{bcsgap0}).
We now assume that the critical temperature is on the order of the 
value of the condensate at zero temperature, $\phi_0$. In analogy to
the treatment for BCS-like theories, we divide
the region of integration into one in which $0 \leq q-\mu \leq \kappa \phi_0$, 
and one in which $\kappa \phi_0 \leq q- \mu \leq \delta$. 
As before, for momenta away from the Fermi surface
the thermal factor is neglected. Using $\phi(y,0)/\phi_0 = \sin(\bar{g}y)$,
we obtain
\begin{equation}
1 \simeq \bar{g}^2 \left[ \int_{x_\kappa}^{x^*} dy \; y\, 
\tanh\left(\frac{\epsilon(y)}{2T}\right) \, \sin(\bar{g}y)
+ \int_{\ln(b\mu/\delta)}^{x_\kappa} dy \; y \, \sin(\bar{g}y)
\right]
\,\, ,
\label{qcdgapT2}
\end{equation}
which is analogous to (\ref{bcsgap000}).
Following the definition of the variable $x$ in (\ref{defvarx}),
here we introduced $x_\kappa = \ln[ b \mu/(\kappa \phi_0)]$.  
Again, in weak coupling 
the $1$ on the left-hand side is almost completely saturated by
the integration region away from the Fermi surface. To see this,
we compute the respective integral in a power series expansion in $\bar{g}$:
\begin{equation} \label{qcdgapT2a}
\bar{g}^2 \int_{\ln(b\mu/\delta)}^{x_\kappa}
dy \; y \, \sin(\bar{g}y) \simeq 1 - \frac{\pi}{2} \, \bar{g} \, \ln (2\kappa)
+ O(\bar{g}^2) \,\, .
\end{equation}
In order to fulfill (\ref{qcdgapT2}), the 
integral over the region very near the Fermi surface has to compensate 
the terms of order $O(\bar{g})$ in Eq.\ (\ref{qcdgapT2a}),
\begin{equation}
\bar{g}^2 \int_{x_\kappa}^{x^*} dy \; y\, 
\tanh\left(\frac{\epsilon(y)}{2T}\right) \, \sin(\bar{g}y)
\simeq \frac{\pi}{2}\, \bar{g}\, \ln(2\kappa) + O(\bar{g}^2) \,\, .
\label{qcdgapT3}
\end{equation}
To see how this compensation works, we expand the various terms on the 
left-hand side in powers of
$\bar{g}$. As $x^*= \pi/(2\bar{g})$, and $x_\kappa = x^* - \ln(2 \kappa)$,
$y$ is of order $x^*$ in the whole integration region, 
up to corrections of order $\bar{g}$. Note that
this is equivalent to approximating the factor 
$\ln(b \mu /\epsilon_q)$ by $\ln (2 b \mu/\phi_0)$.
For the momentum dependence of the gap function, this
has the consequence that $\sin(\bar{g}y) = 1 + O(\bar{g})$, {\it i.e.},
to leading order in weak coupling, the gap
function can be taken to be constant, $ \phi(y,0) \simeq \phi_0$.
In this way one obtains
\begin{equation} \label{qcdgapT4}
\frac{\pi}{2}\, \bar{g} \, \ln (2 \kappa) \simeq
\frac{\pi}{2}\, \bar{g}\, 
\int_0^{\kappa \phi_0} \frac{d(q-\mu)}{\epsilon_q} \, 
\tanh\left(\frac{\epsilon_q}{2T}\right) + O(\bar{g}^2)\,\, ,
\end{equation}
where we reverted the integration variable $y$ into $q-\mu$.
With (\ref{2kappa}), we can write this as
\begin{equation} \label{qcdgapT4a}
0 \simeq \frac{\pi}{2}\, \bar{g} 
\int_0^{\kappa \phi_0} d(q-\mu)\, \left[\, \frac{1}{\epsilon_q} \,  
\tanh \left(\frac{\epsilon_q}{2T}\right) - \frac{1}{\epsilon_q^0} \,
\right] + O(\bar{g}^2) \,\,.
\end{equation}
Apart from the prefactor (which can be divided out),
to leading order in weak coupling,
Eq.\ (\ref{qcdgapT4a}) is the same as in
BCS-type theories, Eq.\ (\ref{bcsgap00})~! 
Thus, also the ratio $\phi(T)/\phi_0$ is {\em identical\/}
to that in BCS.  Given the very different nature of the gap equation,
this is a remarkable result.

How can we claim that we
can reliably compute $\phi(T)/\phi_0$ to leading
order in weak coupling, although we cannot compute
the overall magnitude of the condensate at zero temperature, the
constant $b_0'$ in (\ref{b0})~? 
To understand this, consider the counting of powers of $g$ in the
gap equation; for this, it suffices to use the
approximations of subsection \ref{IIIA}.

{}From Eq.\ (\ref{quadraticeq}) we have seen that
the exponential in $1/g$ arises from terms
$\sim g^2 \ln^2(2 \delta/\phi_0)$. Since $\phi_0 \sim \exp(-1/g)$,
these terms are of order $1$, and therefore of the same order
as the left-hand side of (\ref{quadraticeq}).
Analogously, in the nonzero-temperature gap equation (\ref{qcdgapT2}),
terms of order $1$ arise
from the region of momenta away from the Fermi surface, and
thus balance the $1$ on the left-hand side. As temperature
effects are negligible in this region, we conclude that the exponential
behavior $\sim \exp(-1/g)$ of the gap cannot change with temperature.

The prefactor of the exponential 
is determined by terms $ \sim g^2 \, \ln(2\delta/\phi_0) \sim g$ 
in the gap equation (\ref{quadraticeq}). 
These include {\em all\/} terms of this 
order, {\it i.e.}, the leading logarithmic terms as well as
terms of order one, which give rise to $b_0'$. From Eq.\ (\ref{qcdgapT3})
we realize that temperature effects also enter at order $g$.
These effects therefore change the prefactor of the exponential.
In the above discussion, we have determined the change of the gap 
equation with temperature to leading order in $g$, or in other
words, we have determined the prefactor at nonzero temperature.
To leading order in $g$, the result is that
the prefactor changes precisely in the same way as in BCS-like theories.

To understand how the gap function changes with {\em both\/} $x$ and $T$,
consider first the region just below the critical temperature.
Even though the overall magnitude of the gap, $\phi(T)$, 
is small, we can still consider $\phi(x,T)/\phi(T)$; 
even as $T \rightarrow T_c$, this ratio remains of order one. 
Then near the Fermi surface, the 
gap function {\it must\/} change
due to thermal effects: after all, the variable 
$x^*(T)=\ln[2b\mu/\phi(T)]$ diverges as $T\rightarrow T_c$, when
$\phi(T) \rightarrow 0$.

To understand the change in the gap function, consider
energies small relative to the temperature, 
$\epsilon(x) < T$.  In this limit, 
the thermal factor $\tanh[\epsilon/(2 T)] \simeq \epsilon/(2T)$.
Using the definition of the variable $x$, Eq.\ 
(\ref{diffeq}) is approximately
\begin{equation}
\left(\epsilon \frac{d}{d \epsilon}\right)^2
\phi(\epsilon,T)  \simeq  - \bar{g}^2 \,\left(
\frac{ \epsilon }{2T} \right) \, \phi(\epsilon,T) \,\, .
\label{qcdgapT5}
\end{equation}
About small $\epsilon$, the solution is given as a power series
in $\epsilon/(2T)$:
\begin{equation}
\phi(\epsilon,T) \; \simeq \;
\left[ 1 - \bar{g}^2 \frac{\epsilon}{2T} + O(\bar{g}^4) 
\right] \phi(T) \;\;\; , \;\;\; T \rightarrow T_c
\;\;\; , \;\;\; \epsilon \ll T \; .
\label{qcdgapT6}
\end{equation}
Thus when the energy is much less than the temperature, 
although $x\rightarrow \infty$, the gap function 
is not $\sim \sin(\bar{g}x)$, as for the zero-temperature gap function, 
but approaches a constant.  That is, 
due to the thermal factor in the equation for the gap
function, the momentum dependence is cut off, and the gap function
``flattens''.  This is why in our previous calculation, we could
neglect the momentum dependence of the gap function when the energy
is less than $\kappa \phi_0$. 

\begin{figure} 
\hspace*{2.5cm} 
\psfig{figure=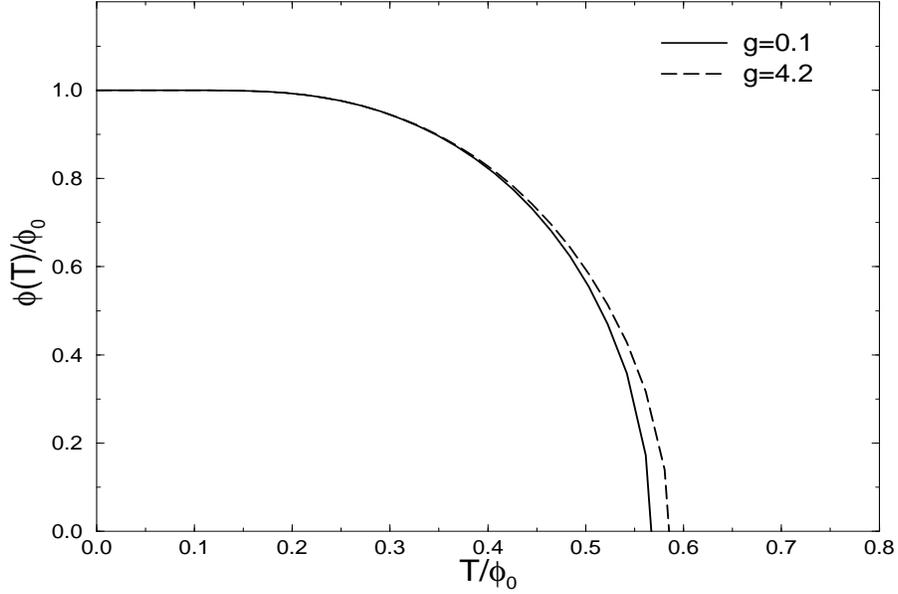,width=3in,height=3.5in,angle=-90}
\vspace*{-1cm}
\caption{The gap at the Fermi surface $\phi(T)$, normalized to its
zero-temperature value $\phi_0$, as a function of $T/\phi_0$.
Solid: $g=0.1$, dashed: $g=4.2$.}
\label{phiofT}
\end{figure}

Our analytical results are confirmed by numerical solutions of 
Eq.\ (\ref{diffeq}) at nonzero temperature.
In Fig.\ \ref{phiofT} we show the gap function at the Fermi surface
as a function of temperature. We considered two values of the coupling
constant, $g=0.1$ and $4.2$. The first is safely in the weak-coupling
regime, while the latter is the value where $\phi_0$ has a maximum
as a function of $g$, cf.\ Fig.\ \ref{phi0}.
In weak coupling, $\phi(T)/\phi_0$ is indistinguishable from BCS.
Surprisingly, even for large $g=4.2$, while $T_c$ is slightly larger
than in BCS theory, $T_c / \phi_0 \simeq 0.585$, $\phi(T)/\phi_0$ is 
rather similar to the behavior in BCS-like theories.
This is not unlike the situation in strongly coupled BCS,
{\it i.e.} Eliashberg theory \cite{Scalapino},
where $T_c/\phi_0$ also changes slightly, albeit in the opposite
direction.

\begin{figure} 
\hspace*{2.5cm} 
\psfig{figure=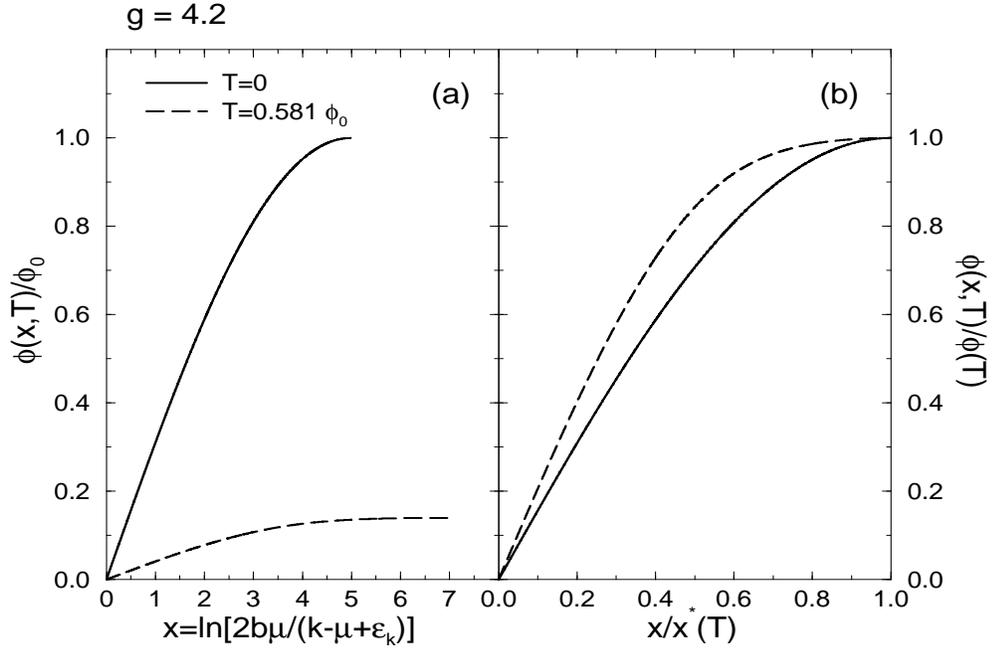,width=3in,height=3.5in,angle=-90}
\vspace*{-1cm}
\caption{(a) The gap function $\phi(x,T)$, normalized to the 
zero-temperature gap function $\phi_0$, as a function of the variable $x$ 
for $T=0$ (solid) and $T=0.581\, \phi_0$ (dashed), $g=4.2$.
(b) The same, but now the gap function is normalized to
its value $\phi(T)$ at the Fermi surface at the same temperature, 
and plotted as function of $x$ normalized to $x^*(T)$.
For the sake of simplicity, we have chosen $\delta = b \mu$.}
\label{gapfunction}
\end{figure}

In Fig.\ \ref{gapfunction} (a) we show $\phi(x,T)$ as a 
function of $x$ at $g=4.2$ for two temperatures, $T=0$ and $T=0.581\, \phi_0$ 
which is close to $T_c$. 
As the temperature increases, the overall
scale of $\phi(x,T)$ decreases, because the condensate is evaporating, cf.\
Fig.\ \ref{phiofT}. Simultaneously, $x^*(T) = \ln [2b\mu/\phi(T)]$ 
increases. In Fig.\ \ref{gapfunction} (b)
we plot $\phi(x,T)/\phi(T)$ as a function of $x/x^*(T)$ at the
same two temperatures. As can be seen, once we divide out the
overall scale $\phi(T)$, the ratio $\phi(x,T)/\phi(T)$ 
is relatively insensitive to temperature.
At $T=0.581\, \phi_0$, we observe that the gap function does ``flatten''
as it approaches the Fermi surface, $x \rightarrow x^*(T)$, as we
argued previously on the basis of Eq.\ (\ref{qcdgapT6}).
\newpage

\vspace*{1cm}

\section{Condensates with total spin $J=1$} \label{J1}

\subsection{Classification} \label{classi}

The gap matrix $\Phi^+$ for
a condensate with total spin $J=0$ has the form (\ref{generalJ0}).
Generalizing this equation to a condensate with total spin $J=1$ is
straightforward. For a $J=1$ condensate,
the individual gap functions $\phi_{hs}^e$
in Eq.\ (\ref{generalJ0}) are no longer scalar functions in 
coordinate space, but 3-vectors, $\bbox{\phi}_{hs}^e$.
These vectors have a component along the direction of ${\bf k}$,
\begin{equation}
\varphi_{hs}^e(K) \equiv \bbox{\phi}_{hs}^e(K) \cdot \hat{\bf k}\,\, ,
\end{equation}
and two components transverse to this direction,
\begin{equation} \label{transverse}
\bbox{\chi}_{hs}^e(K) \equiv \bbox{\phi}_{hs}^e(K) 
\cdot ({\bf 1} - \hat{\bf k}\, \hat{\bf k})\,\, ,
\end{equation}
such that
\begin{equation}
\bbox{\phi}_{hs}^e (K) \equiv \varphi_{hs}^e(K) \, \hat{\bf k}
+ \bbox{\chi}_{hs}^e(K) \,\, .
\end{equation}
The gap matrix $\Phi^+$ is, however, still a scalar in coordinate space.
One therefore has to contract the vector indices of 
$\bbox{\chi}_{hs}^e$ with the other independent 3-vector
at our disposal, $\bbox{\gamma}$,
\begin{equation} \label{contraction}
\bbox{\chi}_{hs}^e (K) \cdot \bbox{\gamma} \equiv
\bbox{\phi}_{hs}^e(K) \cdot ({\bf 1}- \hat{\bf k} \, \hat{\bf k})
\cdot \bbox{\gamma} \equiv \bbox{\chi}_{hs}^e(K) \cdot 
\bbox{\gamma}_\perp ({\bf k}) \,\, ,
\end{equation}
where we have defined $\bbox{\gamma}_\perp({\bf k}) \equiv 
({\bf 1} - \hat{\bf k} \, \hat{\bf k}) \cdot \bbox{\gamma}$ and
used the projector property $({\bf 1} - \hat{\bf k} \, \hat{\bf k})^2
\equiv {\bf 1} - \hat{\bf k} \, \hat{\bf k}$.

Suppressing color and flavor indices for the moment, the most
general ansatz for a gap matrix describing $J=1$ condensates is therefore
\begin{equation} \label{generalJ1}
J=1:\;\;\;\;
\Phi^+(K) = \sum_{h=r,\ell}\; \sum_{s=\pm}\; \sum_{e=\pm}
\left[\varphi_{hs}^e(K) + \bbox{\chi}_{hs}^e(K)
     \cdot \bbox{\gamma}_\perp ({\bf k}) \right] \, 
{\cal P}_{hs}^e({\bf k}) \,\, . 
\end{equation}
The scalar product $\bbox{\chi}_{hs}^e(K) \cdot
\bbox{\gamma}_\perp ({\bf k})$ has simple commutation properties with the
quasiprojectors ${\cal P}_{hs}^e({\bf k})$:
\begin{mathletters} \label{commutations}
\begin{eqnarray}
{\cal P}_h\,  \bbox{\gamma}_\perp({\bf k}) & = & 
\bbox{\gamma}_\perp({\bf k}) \,{\cal P}_{-h} \,\, , \\
{\cal P}_{s} ({\bf k}) \, \bbox{\gamma}_\perp({\bf k}) & = & 
\bbox{\gamma}_\perp({\bf k}) \, {\cal P}_{-s} ({\bf k}) \,\, , \\
\Lambda^e ({\bf k}) \, \bbox{\gamma}_\perp({\bf k}) & = & 
\bbox{\gamma}_\perp({\bf k}) \, \left[\Lambda^e ({\bf k})
-e\, \alpha_{\bf k} \gamma_0 \right]\,\, . \label{commutations3}
\end{eqnarray}
\end{mathletters}
In the massless limit, $\alpha_{\bf k} =0$, and $\bbox{\gamma}_\perp$
commutes with the energy projectors.

Instead of contracting the {\em transverse\/} components (\ref{transverse})
of $\bbox{\phi}_{hs}^e(K)$ with $\bbox{\gamma}$, one could have alternatively
also used $\bbox{\phi}_{hs}^e(K)$ itself. The commutation properties
for $\bbox{\phi}_{hs}^e \cdot \bbox{\gamma}$, 
however, are more complicated than for
$\bbox{\chi}_{hs}^e \cdot \bbox{\gamma}_\perp({\bf k})$, with
additional terms proportional to $\hat{\bf k}$. 
The advantage of using $\bbox{\gamma}_\perp({\bf k})$ is that these 
terms are cancelled by the projector ${\bf 1} -
\hat{\bf k} \, \hat{\bf k}$.

Finally, an important question that arises is why we choose
$\bbox{\gamma}$ to contract the transverse components
of $\bbox{\phi}_{hs}^e$ in Eq.\ (\ref{generalJ1}). 
Why not $\gamma_0\, \bbox{\gamma}$,
or $\gamma_5\, \gamma_0\, \bbox{\gamma}$~?
The answer lies again in the commutation properties
(\ref{commutations}).
In the commutation property for $\gamma_0\, \bbox{\gamma}_\perp({\bf k})$ 
or $\gamma_5 \, \gamma_0\, \bbox{\gamma}_\perp({\bf k})$
corresponding to (\ref{commutations3}), 
the sign of the energy on the right-hand side would be flipped.
Physically, this means that such terms represent pairing
of particles with antiparticles, and not of particles with
particles, cf.\ discussion of
Eq.\ (\ref{decomposition}) in subsection \ref{ssurlimit}.
This is clearly not what happens in a superconductor, where
particles form pairs near the Fermi surface.

\subsection{Parity} \label{parity}

The $J=1$ condensates can be decomposed according to their
transformation properties under parity.
Let us first write the interaction term $\bar{\psi}_C\, \Phi^+\, \psi$
in the effective action \cite{prscalar} in the form
\begin{equation} \label{interaction}
\bar{\psi}_C(K)\, \Phi^+(K)\, \psi(K) = 
\sum_{h=r,\ell}\; \sum_{s=\pm}\; \sum_{e=\pm} \bar{\psi}_C(K)\, 
\bbox{\phi}_{hs}^e(K) \cdot \left( \hat{\bf k} + \bbox{\gamma}_\perp
({\bf k}) \right)\, {\cal P}_{hs}^e ({\bf k}) \,\psi(K)\,\, .
\end{equation}
Let us define new condensate fields
\begin{equation}
\bbox{\phi}^+_\pm \equiv \frac{1}{2} \left( 
\bbox{\phi}_{r+}^+ \pm \bbox{\phi}_{\ell-}^+ \right) 
\;\; , \;\;\;
\bbox{\phi}^-_\pm \equiv \frac{1}{2} \left( 
\bbox{\phi}_{r-}^- \pm \bbox{\phi}_{\ell+}^- \right) 
\;\; , \;\;\;
\bbox{\pi}^+_\pm \equiv \frac{1}{2} \left( 
\bbox{\phi}_{r-}^+ \pm \bbox{\phi}_{\ell+}^+ \right) 
\;\; , \;\;\;
\bbox{\pi}^-_\pm \equiv \frac{1}{2} \left( 
\bbox{\phi}_{r+}^- \pm \bbox{\phi}_{\ell-}^- \right) 
\,\, .
\end{equation}
In terms of these fields, the interaction term (\ref{interaction})
assumes the form
\begin{eqnarray}
\lefteqn{ \bar{\psi}_C(K)\, \Phi^+(K)\, \psi(K)  } \nonumber \\
& = & \sum_{e=\pm} \bar{\psi}_C(K)\, \left\{ \left[
  \bbox{\phi}_+^e(K) \cdot \left( \hat{\bf k} + \bbox{\gamma}_\perp
({\bf k}) \right)
+ \bbox{\phi}_-^e(K) \cdot \left(\hat{\bf k} + \bbox{\gamma}_\perp
({\bf k}) \right) \gamma_5 \, \right]
\frac{1+e\,\gamma_0\, \bbox{\gamma} \cdot \hat{\bf k}}{2}
\, \Lambda^e ({\bf k}) \right. \nonumber \\
&   & \hspace*{1.7cm} \left. + \left[
  \bbox{\pi}_+^e(K) \cdot \left( \hat{\bf k} + \bbox{\gamma}_\perp
({\bf k}) \right)
+ \bbox{\pi}_-^e(K) \cdot \left(\hat{\bf k} + \bbox{\gamma}_\perp
({\bf k}) \right) \gamma_5 \, \right]
\frac{1-e\,\gamma_0\, \bbox{\gamma} \cdot \hat{\bf k}}{2}
\, \Lambda^e ({\bf k}) \right\} \psi(K) \,\, . \label{int2}
\end{eqnarray}
In the limit $m\rightarrow 0$, $\Lambda^e({\bf k}) \rightarrow 
(1+e\,\gamma_0\, \bbox{\gamma} \cdot \hat{\bf k})/2$, such that
the terms proportional to $\bbox{\pi}^e_\pm$ vanish.

Under a parity transformation, $\gamma_0 \rightarrow \gamma_0$,
$\bbox{\gamma} \rightarrow - \bbox{\gamma}$, 
$\gamma_5 \rightarrow - \gamma_5$, and ${\bf k} \rightarrow - {\bf k}$.
Thus, the terms $(1\pm e\,\gamma_0\, \bbox{\gamma} \cdot \hat{\bf k})/2$
and the energy projectors do not change under
a parity transformation. The parity of $\bar{\psi}_C$ is opposite to
that of $\psi$ \cite{prscalar}. A parity transformation leaves
the effective action invariant, since the QCD Lagrangian (from which the
effective action is derived) does not violate parity explicitly.
Thus, if we perform a parity transformation of the interaction 
term (\ref{int2}), the gap functions $\bbox{\phi}^e_\pm, \, 
\bbox{\pi}^e_\pm$ must have the same parity as the accompanying 
Dirac matrices. Consequently, we conclude
that $\bbox{\phi}_+^e,\, \bbox{\pi}_+^e$ are
parity-even, while $\bbox{\phi}_-^e,\, \bbox{\pi}_-^e$ are parity-odd.

Two limiting cases are of special interest. If all right-handed, 
positive-helicity gaps are equal to the left-handed, negative-helicity gaps, 
$\bbox{\phi}_{r +}^e = \bbox{\phi}_{\ell -}^e$, and all
right-handed, negative-helicity gaps are equal to 
the left-handed, positive-helicity
gaps, $\bbox{\phi}_{r -}^e = \bbox{\phi}_{\ell +}^e$, then
$\bbox{\phi}^e_- = \bbox{\pi}_-^e =0$, {\it i.e.}, the
odd-parity gaps vanish, and condensation
occurs exclusively in the $J^P = 1^+$ channel. [It is proven in 
Appendix \ref{A} that in this case Eq.\ (\ref{int2}) agrees with
the ansatz of Bailin and Love for the $J^P=1^+$ gaps \cite{bailinlove}.]
On the other hand, if they are equal in magnitude, but 
different in sign, condensation occurs exclusively in the $J^P=1^-$ channel.

This is different for the $J=0$ gaps. In that case, one can write 
down an equation analogous to (\ref{int2}). The difference is that
all vectors $\bbox{\phi}_\pm^e, \, \bbox{\pi}_\pm^e$ are replaced by
scalar functions $\phi_\pm^e,\, \pi_\pm^e$, and the terms
$\hat{\bf k} + \bbox{\gamma}_\perp({\bf k})$ are absent, too.
For the parity transformation properties, this has the consequence
that if $\phi_{r +}^e =\pm \phi_{\ell -}^e$, 
$\phi_{r -}^e =\pm \phi_{\ell +}^e$, condensation occurs in the
$J^P = 0^\mp$ channel.

\subsection{The ultrarelativistic limit} \label{ssurlimit}

In the following, we exclusively consider massless fermions.
In this case, the helicity projectors (and corresponding
indices) become superfluous, cf.\ Eq.\ (\ref{urlimit}),
\begin{equation}
m=0:\;\;\;\;\varphi_{hs}^e \longrightarrow \varphi_h^e \;\;\;\; , 
\;\;\;\; \bbox{\chi}_{hs}^e \longrightarrow \bbox{\chi}_h^e \,\, ,
\end{equation}
and the quasiprojectors become true projectors, cf.\ discussion
above [and Eq.\ (B29) \cite{prscalar}].
The general ansatz (\ref{generalJ1}) simplifies to
\begin{equation}
J=1,\, m=0:\;\;\;\;
\Phi^+(K) = \sum_{h=r,\ell}\; \sum_{e=\pm}
\left[\varphi_h^e(K) + \bbox{\chi}_h^e(K) \cdot \bbox{\gamma}_\perp 
({\bf k}) \right] \, {\cal P}_h^e ({\bf k}) \,\, .
\end{equation}
In the effective action, the interaction term $\bar{\psi}_C \, \Phi^+\, \psi$
decomposes into four parts on account of 
the projectors ${\cal P}_h^e$ [cf.\ Eq.\ (B34) of \cite{prscalar}],
\begin{equation} \label{decomposition}
\bar{\psi}_C(K) \, \Phi^+(K) \, \psi(K) = \sum_{h=r,\ell}\; 
\sum_{e=\pm} \left[ \bar{\psi}_{C\,h}^e(K) \, \varphi_h^e(K) \, \psi_h^e(K)
+ \bar{\psi}_{C\,-h}^e(K) \, \bbox{\chi}_h^e(K) \cdot \bbox{\gamma}_\perp 
({\bf k})\, \psi_h^e(K) \right]\,\,.
\end{equation}
Here we defined $\psi_h^e \equiv {\cal P}_h^e \, \psi$,
and used the commutation properties (\ref{commutations}) of
$\bbox{\gamma}_\perp$. From (\ref{decomposition}) it is obvious
that the longitudinal gaps correspond to condensation
of quarks with the {\em same\/} chirality, $(rr)$ or $(\ell \ell)$.
In this respect the longitudinal $J=1$ gaps are similar to the
$J=0$ gaps \cite{prscalar}.
On the other hand, the transverse $J=1$ gaps correspond to condensation of
quarks with {\em different\/} chiralities, $(r\ell)$ or
$(\ell r)$.

Equation (\ref{decomposition}) shows why we do not choose
$\gamma_0 \, \bbox{\gamma}$ or $\gamma_5\,
\gamma_0\, \bbox{\gamma}$ to contract the transverse components
of $\bbox{\phi}_{hs}^e$ in (\ref{generalJ1}), cf.\ discussion at
the end of subsection \ref{classi}. The extra factors
$\gamma_0$ or $\gamma_5\, \gamma_0$ flip the sign of
the energy in (\ref{commutations3}), 
and the transverse gaps would describe pairing of particles with 
antiparticles of the same chirality. 

\subsection{Color and flavor representations}

With the help of the symmetry property [cf.\ Eq.\ (B4) of \cite{prscalar}]
\begin{equation}
C \, \Phi^+(K) C^{-1} = \left[\Phi^+(-K)\right]^T\,\, ,
\end{equation}
we derive
\begin{equation} \label{symmetryprops}
\left[\varphi_h^e(-K)\right]^T = - \varphi_h^e (K)
\;\;\;\; , \;\;\;\;
\left[\bbox{\chi}_h^e(-K)\right]^T = - \bbox{\chi}_{-h}^e(K)
\,\, .
\end{equation}
The symmetry properties (\ref{symmetryprops}) allow us to
classify the possible color and flavor representations of
the $J=1$ condensates for massless fermions. 
We assume there are $N_f$ flavors of massless fermions with a
global flavor symmetry $SU(N_f)_r \times SU(N_f)_\ell$ and,
of course, a local color symmetry $SU(3)_c$. From group theory,
\begin{equation}
{\bf 2} \times {\bf 2} = {\bf 1}_a + {\bf 3}_s \;\;\;\; , \;\;\;\;
{\bf 3} \times {\bf 3} = \bar{\bf 3}_a + {\bf 6}_s \,\, ,
\end{equation}
where the subscripts $a$ and $s$ denote antisymmetric or
symmetric representations, respectively.
For single-gluon exchange, the color-antitriplet channel $\bar{\bf 3}^c_a$ 
is attractive, and the color-sextet channel ${\bf 6}^c_s$ is repulsive.

We first discuss the longitudinal condensates $\varphi_h^e$. From 
Eq.\ (\ref{decomposition}), the longitudinal $J=1$ condensates 
only couple quarks of the same chirality, and so do not break
the $SU(N_f)_r \times SU(N_f)_\ell$ flavor symmetry. 
By (\ref{symmetryprops}), the longitudinal condensates must
correspond to a color-flavor representation which is overall
antisymmetric. 

For $N_f=1$, the flavor representation is trivial, and
condensation must occur in the $\bar{\bf 3}^c_a$ channel.
This is most likely the favored channel for condensation of
quarks of the same flavor, since the $J=0$ gaps are overall 
symmetric \cite{prlett,prscalar}, and consequently must be in the repulsive 
${\bf 6}^c_s$ channel.

For $N_f =2$, the allowed color-flavor representations are either
$(\bar{\bf 3}^c_a,\, {\bf 3}^f_s)$, which is favored, or 
$({\bf 6}^c_s,\, {\bf 1}^f_a)$. 
This is in contrast to the $J=0$ gaps, which come in
$(\bar{\bf 3}^c_a,\, {\bf 1}^f_a)$ or $({\bf 6}^c_s, \, {\bf 3}^f_s)$.
Here the flavor representation refers
to either $SU(2)_r$ or $SU(2)_\ell$.

For $N_f=3$, there are
$(\bar{\bf 3}^c_a,\, {\bf 6}^f_s)$ or $({\bf 6}^c_s,\, \bar{\bf 3}^f_a)$.
On the other hand, the $J=0$ gaps are
$(\bar{\bf 3}^c_a,\, \bar{\bf 3}^f_a)$ or $({\bf 6}^c_s, \, {\bf 6}^f_s)$.

If only longitudinal $J=1$ gaps are present, the parity properties are
analogous to the $J=0$ gaps. In particular, since condensates of
different chiralities do not mix, the magnitude of the longitudinal
$J=1$ gaps will be equal, while their relative phase represents the
spontaneous breaking of parity \cite{prlett,parity}.

The symmetry relation (\ref{symmetryprops}) relates transverse
$J=1$ condensates of different chirality.
Therefore, we cannot draw general conclusions about the
possible color-flavor representations of the transverse $J=1$ gaps.
{\em A priori\/}, both the symmetric as well as the antisymmetric 
color and flavor representations are allowed.

Two special cases are of interest.
If $\bbox{\chi}_{r,\ell}^e \equiv \bbox{\chi}_{\ell,r}^e$, so that condensation
is in the $J^P=1^+$ channel, the
color-flavor representations of the transverse $J=1$ gaps are
identical to those of the longitudinal $J=1$ gaps. 
On the other hand, if $\bbox{\chi}_{r,\ell}^e \equiv - \bbox{\chi}_{\ell,r}^e$,
so that condensation is in the $J^P=1^-$ channel,
they are equal to those of the $J=0$ gaps.

When $N_f \geq 2$, the appearance of transverse $J=1$ gaps must
necessarily break the $SU(N_f)_r \times SU(N_f)_\ell$ symmetry to
a vector-like $SU(N_f)$ symmetry. This striking feature arises because
the transverse gaps are proportional to $\bbox{\gamma}_\perp$, and not,
say, $\gamma_0 \, \bbox{\gamma}_\perp$.

\subsection{The gap equation}

In general, condensation can occur in channels with arbitrary
total spin $J$. Therefore, the gap matrix $\Phi^+(K)$ will not simply be 
of the $J=0$ form (\ref{generalJ0}), or the $J=1$ form (\ref{generalJ1}),
but it will be a sum of (\ref{generalJ0}) and (\ref{generalJ1}), as
well as contain contributions from higher spin $J \geq 2$.
We do not attempt to solve this problem in full generality. Instead, we
consider the simpler case where the gap matrix contains
just $J=0$ and $J=1$ contributions. We also take the fermions to be massless.
The gap matrix then assumes the form 
\begin{equation}
J=0\; {\rm and} \; 1,\;\; m=0:\;\;\;\;
\Phi^+(K) = \sum_{h=r,\ell}\; \sum_{e=\pm}
\left[\phi_h^e(K) + \varphi_h^e(K) + \bbox{\chi}_h^e(K) 
\cdot \bbox{\gamma}_\perp ({\bf k})\right] \, {\cal P}_h^e({\bf k}) \,\, . 
\end{equation}
where $\phi_h^e$ denote the $J=0$ gaps, and $\varphi_h^e$
the longitudinal and $\bbox{\chi}_h^e$ the transverse
components of the $J=1$ gaps.

The quasiparticle propagator can be computed from Eqs.\ (\ref{quasiprop}),
(\ref{selfenergy}), and (\ref{chargeconjcond}), and the
commutation property (\ref{commutations3}).
Ignoring the color and flavor structure for the moment, one obtains
\begin{eqnarray} \label{quasipropJ01}
\lefteqn{G^+(Q) =
\left\{\sum_{h=r,\ell}\; \sum_{e=\pm}
\left[\frac{}{} q_0^2 - (q-e\mu)^2 \right. \right.}  \\ 
& + & \left. \left. \left( \frac{}{} [\phi_h^e(Q)]^\dagger 
+[\varphi_h^e(Q)]^\dagger - [\bbox{\chi}_{-h}^e(Q)]^\dagger \cdot 
\bbox{\gamma}_\perp ({\bf q}) \right) \left( \frac{}{} \phi_h^e(Q)
+\varphi_h^e(Q) + \bbox{\chi}_h^e(Q) \cdot \bbox{\gamma}_\perp
({\bf q}) \right)
\right] {\cal P}_h^e({\bf q}) \right\}^{-1} \left[G_0^-(Q) \right]^{-1}.
\nonumber
\end{eqnarray}
To invert the term in curly brackets, we first have to
specify the color-flavor structure of the various condensates.
In the following, we just consider the most simple case $N_f=1$.
Let us assume that condensation occurs exclusively in the
attractive color-antitriplet channel $\bar{\bf 3}^c_a$, {\it i.e.},
$\Phi^+_{ij} \equiv \epsilon_{ijk} \Phi^+_k$ is an antisymmetric
$N_c \times N_c$ matrix in color space, $i,j= 1, \ldots, N_c$.
In this case, the $J=0$ gaps vanish, $\phi_h^e \equiv 0$, 
on account of Fermi statistics \cite{prlett,prscalar}.
Since the individual gap functions $\varphi_{h\,ij}^e, \,
\bbox{\chi}_{h\, ij}^e$ are also in the $\bar{\bf 3}^c_a$
representation, we conclude from (\ref{symmetryprops}) that
$\bbox{\chi}_r^e \equiv \bbox{\chi}_\ell^e$.
In contrast, $\varphi_r^e$ and $\varphi_\ell^e$ remain {\em
a priori\/} unrelated.

The term in curly brackets in (\ref{quasipropJ01}) has an off-diagonal
contribution in color space. Since its contribution to
$G^+_{ij}$ is quadratic in 
$\bbox{\phi}_h^e$, we shall neglect it in the following, and
consider only the diagonal part, $G^+_{ij} \simeq \delta_{ij}\,
G^+$. Even so, the inversion of the term in curly brackets is still
cumbersome, due to the presence of terms $\sim \bbox{\gamma}_\perp$.
To simplify the treatment, we assume that the
$J=1$ gaps are {\em real-valued}.
With Eq.\ (\ref{symmetryprops}) this leads to
\begin{equation}
\left[ \varphi_h^e (K) \right]^\dagger  = 
\left[ \varphi_h^e (-K) \right]^T
\equiv - \varphi_h^e (K) 
\;\;\;\; , \;\;\;\;
\left[ \bbox{\chi}_{-h}^e (K) \right]^\dagger  = 
\left[ \bbox{\chi}_{-h}^e (-K) \right]^T
\equiv - \bbox{\chi}_h^e (K) \,\, .
\end{equation}
This has the consequence that all cross terms $\sim \bbox{\gamma}_\perp$
between the gaps $\varphi_h^e$ and $\bbox{\chi}_h^e$ vanish
in (\ref{quasipropJ01}).
The quasiparticle propagator can now be explicitly computed:
\begin{equation}
G^+ (Q)  \simeq  \sum_{h=r,\ell}\; \sum_{e=\pm}
\frac{{\cal P}_h^e({\bf q})}{q_0^2  - 
\left[\epsilon_q^e(\bbox{\phi}_h^e) \right]^2}\, 
\left[G_0^- (Q) \right]^{-1} \,\, . \label{prop}
\end{equation}
The quasiparticle energies are
\begin{equation} \label{quasipenergy}
\epsilon_q^e (\bbox{\phi}_h^e) \equiv 
\left[ (q - e\, \mu)^2 + \sum_{l= 1}^{N_c} \bbox{\phi}_{h\, l}^e
\cdot \bbox{\phi}_{h\, l}^e \right]^{1/2} \,\, .
\end{equation}
Inserting (\ref{prop}) into the gap equation (\ref{gapeq}), the same
steps that led to (\ref{19}) now lead to
\begin{equation}
\Phi^+_i(K) \simeq \frac{2}{3}\, g^2 \frac{T}{V} \sum_Q
\gamma^\mu\, \Delta_{\mu \nu}(K-Q) 
\sum_{h=r,\ell} \;\sum_{e=\pm} \frac{\varphi_{h\, i}^e(Q)-
\bbox{\chi}_{h\, i}^e(Q)\cdot \bbox{\gamma}_\perp({\bf q})}{
q_0^2 - \left[\epsilon^e_q(\bbox{\phi}_h^e)\right]^2}  
\, {\cal P}_{-h}^{-e}({\bf q})\,  \gamma^\nu \,\, .
\end{equation}
The gap equations for different fundamental colors
$i$ decouple, therefore we omit the color index in the following.
Taking projections we arrive at the following equation for
the longitudinal gap functions:
\begin{eqnarray}\label{projectedgapeqJ1+}
\varphi_h^e(K) & \simeq & \frac{2}{3}\, g^2 \frac{T}{V} \sum_Q
\Delta_{\mu \nu}(K-Q) \, \left\{
\frac{\varphi_h^e(Q)}{q_0^2 - 
\left[\epsilon^e_q(\bbox{\phi}_h^e)\right]^2}  
 \, {\rm Tr} \left[{\cal P}_h^e({\bf k})\, \gamma^\mu \,
{\cal P}_{-h}^{-e}({\bf q})\, \gamma^\nu \right] \right. \nonumber \\
&   & \hspace*{3.6cm} \left.
+\, \frac{\varphi_h^{-e}(Q)}{q_0^2- 
\left[\epsilon^{-e}_q(\bbox{\phi}_h^{-e})\right]^2} 
 \, {\rm Tr} \left[{\cal P}_h^e({\bf k}) \, \gamma^\mu\, 
{\cal P}_{-h}^e({\bf q}) \, \gamma^\nu \right] \right\} \,\, .
\end{eqnarray}
This equation is rather similar
to Eq.\ (\ref{projectedgapeq}). The only difference
is the appearance of the transverse gaps in the quasiparticle and
quasi-antiparticle energies $\epsilon^e_q(\bbox{\phi}_h^e)$, cf.\
Eq.\ (\ref{quasipenergy}). 

For the transverse gaps, we use the fact that
$\bbox{\chi}_r^e = \bbox{\chi}_\ell^e$ to arrive at
\begin{eqnarray}
\bbox{\chi}_h^e (K) & \simeq & \frac{2}{3}\, g^2 \frac{T}{V} \sum_Q
\Delta_{\mu \nu}(K-Q) \, \left\{ \frac{\bbox{\chi}_h^e(Q)}{q_0^2 -
 \left[\epsilon^e_q(\bbox{\phi}_h^e)\right]^2}  \cdot
 \frac{1}{2}\, 
{\rm Tr} \left[\bbox{\gamma}\, \Lambda^{-e}({\bf q}) \, \gamma^\nu \,
\Lambda^e({\bf k})\, \bbox{\gamma}_\perp({\bf k})\, \gamma^\mu \right] 
\right. \nonumber \\
&   & \hspace*{3.6cm} \left.
+\, \frac{\bbox{\chi}_h^{-e}(Q)}{q_0^2 -
 \left[\epsilon^{-e}_q(\bbox{\phi}_h^{-e})\right]^2} \cdot  
 \frac{1}{2}\, 
{\rm Tr} \left[\bbox{\gamma}\, \Lambda^e({\bf q}) \, \gamma^\nu \,
\Lambda^e({\bf k})\, \bbox{\gamma}_\perp({\bf k}) \, \gamma^\mu  \right] 
\right\} \,\, .
\end{eqnarray}
Taking Coulomb gauge for the gluon propagator, Eq.\
(\ref{Coulombgauge}), and computing the traces similar to Eqs.\ (\ref{traces}),
we obtain
\begin{eqnarray}
\bbox{\chi}_h^e(K) & \simeq & \frac{2}{3}\, g^2 \frac{T}{V} \sum_Q \left\{
\frac{\bbox{\chi}_h^e(Q)}{q_0^2 - 
\left[\epsilon^e_q(\bbox{\phi}_h^e)\right]^2}  \cdot
\left( {\bf 1}\, \frac{1 + \hat{\bf k} \cdot \hat{\bf q}}{2}
- \hat{\bf k}\, \frac{\hat{\bf k} + \hat{\bf q}}{2} \right) 
\right. \nonumber \\
&   & \hspace*{4.3cm} \times \left[ \Delta_l(K-Q) - \Delta_t(K-Q)\,
\left(1- \frac{(k- q)^2}{({\bf k} - {\bf q})^2} \right)
\right] \nonumber \\
&   & \hspace*{1.6cm} \left.
+ \frac{\bbox{\chi}_h^{-e}(Q)}{q_0^2 - 
\left[\epsilon^{-e}_q(\bbox{\phi}_h^{-e})\right]^2}  \cdot
\left( {\bf 1}\, \frac{1 - \hat{\bf k} \cdot \hat{\bf q}}{2}
- \hat{\bf k}\, \frac{\hat{\bf k} - \hat{\bf q}}{2} \right) 
\right. \nonumber \\ 
&   & \hspace*{4.3cm} \left. \times \left[ \Delta_l(K-Q) - \Delta_t(K-Q)\,
\left(1- \frac{(k+ q)^2}{({\bf k} - {\bf q})^2}\right) \right]
\right\} \,\, .
\end{eqnarray}
Neglecting the antiparticle contribution,
and taking $k = q = \mu$, as well as ${\bf k} \simeq {\bf q}$
(the collinear limit; in Eq.\ (\ref{particlegap2}) this corresponds
to neglecting terms of order $p^2$),
we obtain for the transverse quasiparticle gap 
\begin{equation}
\bbox{\chi}_h^+ (K) \simeq \frac{2}{3}\, g^2 \frac{T}{V} \sum_Q 
\frac{\bbox{\chi}_h^+(Q)}{q_0^2 - 
\left[\epsilon^+_q(\bbox{\phi}_h^{+})\right]^2} \, 
\left[ \Delta_l(K-Q) - \Delta_t(K-Q) \right] \,\, .
\end{equation}
To leading logarithmic order, the gap equation for the transverse gap is 
{\em identical\/} to that for the longitudinal gap. 
To see this directly, one computes the
traces in Eq.\ (\ref{projectedgapeqJ1+}).
One obtains an equation very similar to Eq.\ (\ref{projectedgapeq2}).
Now one computes the angular factors 
for ${\bf k} \simeq {\bf q}$, $k=q  = \mu$:
\begin{equation}
\frac{1+\hat{\bf k} \cdot \hat{\bf q}}{2} \simeq 1\;\;\; , 
\;\;\;\;
- \frac{3-\hat{\bf k} \cdot \hat{\bf q}}{2} + 
\frac{1+\hat{\bf k} \cdot \hat{\bf q}}{2} \, \frac{(k-q)^2}{
({\bf k} - {\bf q})^2} \simeq - 1 \,\,,
\end{equation} 
which proves our assertion.

Consequently, to
this order all $J=1$ gaps are of equal magnitude. Further,
the sum of the squares of the $J=1$ gaps fulfills the same gap
equation as the square of the $J=0$ gap. This conclusion agrees
with Son's renormalization-group analysis \cite{Son}, who argued that
the parametric dependence on $g$ of any spin $J$ gap is the same.
We find that even the prefactor is the same to leading logarithmic
accuracy. Our results differ from those of Hsu and Schwetz \cite{hsuschwetz},
who argued that $J=0$ gaps are favored over those for higher spin.

Beyond leading logarithmic order we suggest that, if 
$J=0$ gaps are allowed by color-flavor symmetry, they are
probably favored over the $J=1$ gaps. If only a $J=1$ gap is allowed, 
as for $N_f =1$, we believe that either the longitudinal gaps, or 
the transverse gaps with a definite color-flavor representation, will be 
favored. Which one is favored will be determined by the constants
$b_0'$ and $b_1'$ in Eqs.\ (\ref{b0}) and (\ref{phi1}).

\section{Conclusions} \label{V}

We conclude by addressing effects which can contribute
to the constants $b_0'$ and $b_1'$ in the condensate, Eqs.\
(\ref{b0}) and (\ref{phi1}). We suggest that our lengthy calculations which
led to the result (\ref{b0}) for $b_0$ may be done much more
efficiently by constructing an effective theory for quarks near 
the Fermi surface, as initiated by Hong \cite{hong}. 
This is presumably the easiest way
to calculate $b_0'$ and $b_1'$ as well. Nevertheless, we
can estimate what kind of effects could contribute to these as of yet
undetermined constants. These include:

{\bf (i) Gauge dependence of the condensate:} \\
If calculated properly, any physical quantity must be independent of 
the choice of gauge.
For color superconductivity, what is physical is the gap on the
quasiparticle mass shell.  At nonzero
temperature and zero quark density, general arguments due to 
Kobes, Kunstatter, and Rebhan \cite{kkr} indicate that the mass shells 
for quarks and gluons are gauge invariant.  Their proof does not extend
obviously to nonzero density, but we shall assume that to be
the case.  

At higher orders, the quasiparticle self energy, $\Sigma^+$, 
contains not only the interaction with the condensate, 
Eq.\ (\ref{selfenergy}), but also
wave-function renormalization. Any apparent gauge dependence
in the former must be cancelled by the gauge dependence of
wave-function renormalization, leaving the quasiparticle mass shell
gauge invariant.

It is easy to see that such subtleties do not enter at the order
to which we have computed.  The gluon propagator in Coulomb gauge,
with gauge fixing parameter $\xi_C$, is given in (\ref{Coulombgauge}).
Alternatively, one could have taken covariant
gauge; with gauge fixing parameter $\xi$, the covariant gauge
propagator was given in Eq.\ (37) of \cite{RDPphysica}.
We have seen, however, that to the accuracy to which we compute, only
the gluon propagator in the static limit, $p_0 \rightarrow 0$, matters.
In the static limit, the gauge dependent terms are identical for
either Coulomb or covariant gauges, and appear only as spatially
longitudinal terms, $\sim \xi\, \hat{p}^i \,\hat{p}^j/p^2$.  
Consider, for example, how the gauge dependent part of the gluon
propagator affects the
quasiparticle contribution, $\sim \phi_h^+/\left\{q_0^2- 
\left[\epsilon^+_q(\phi_h^+)\right]^2\right\} $,
to the gap equation for $\phi_h^+$. From the gap equation 
(\ref{projectedgapeq}), and using (\ref{traces}), this becomes
\begin{equation}
\xi_C \; {\rm Tr} \left[ {\cal P}_h^+({\bf k}) \, \bbox{\gamma}
\cdot \hat{\bf p} \, {\cal P}_{-h}^- \, ({\bf q})
\, \bbox{\gamma} \cdot \hat{\bf p}\right] \sim 
- \xi_C \, \frac{1 + \hat{\bf k} \cdot \hat{\bf q}}{2}\, 
\frac{(k - q)^2}{p^2} \rightarrow 0 \;\; , \;\; k,q \rightarrow \mu \; .
\end{equation}
That is, there are gauge dependent terms, but 
they only contribute to the antiparticle gaps, $\phi_h^-$.
(Further, the antiparticle gaps must be computed on 
their proper mass shell. At the Fermi surface, $\epsilon_k^-\sim 2\, \mu$ 
is not small, in contrast to $\epsilon_k^+ \sim \phi_0 \ll \mu$.)
Consequently, gauge dependent terms in the particle gaps 
do not appear to even affect the prefactor in the condensate,
the constant $b_0'$. These conclusions about 
gauge invariance agree with the results
of Sch\"afer and Wilczek \cite{SchaferWilczek}.  In contrast, 
Hong {\it et al.} \cite{hongetal} argue that Landau gauge is preferred, as in
other approximate treatments of Schwinger-Dyson equations.  We
insist that in the present example, direct calculation demonstrates
gauge invariance without further ado.

{\bf (ii) Wave-function and vertex renormalization:}\\  As noted by
Son \cite{Son}, one can have infrared singular factors for wave-function
renormalization.  For non-relativistic fermions,
this was noted long ago by Holstein, Norton,
and Pincus \cite{holstein}.  (This wave-function renormalization
is not the HDL correction discussed by Sch\"afer
and Wilczek \cite{SchaferWilczek}; such corrections involve
two hard lines, and are down by $g^2$.)  
The dominant corrections involve
a very soft transverse gluon on a quark line; this produces 
gauge-invariant terms of the form $Z - 1 \sim g^2 \ln(\mu/\epsilon_q)
\sim g $ when $\epsilon_q \sim \phi_0$. 
This correction was computed by Brown, Liu, and Ren \cite{rockefeller}, who
find that it is a large effect.  
Besides such wave-function renormalization, one might expect
that the Slavnov--Taylor identities would also generate similar
corrections for the gluon, and for the quark-quark-gluon vertex.
This was not found, however, by the authors of \cite{rockefeller}.

{\bf (iii) Effects of the condensate:}\\ We have computed the gap using
an HDL-resummed gluon propagator.  This is possible because the momenta
which generate the gap are much larger than the scale of
the gap.  To see this, note that in the transverse gluon propagator, 
(\ref{approxspec}),
Landau damping contributes to the gluon propagator when
the momentum $p^6 \sim (m_g^2 \omega)^2$; since
the frequency $\omega \sim \phi$, the dominant momenta are
$p \sim m_g^{2/3} \phi^{1/3}$, which for small $\phi$ is much
larger than $\phi$.  

Effects of the condensate on the gluon propagator can be
estimated by power counting at large momentum, $p \gg \phi_0$.  One would
naturally expect that they are $\sim g^2 \phi^2$, but due to
an infrared divergence, they are larger, 
$\sim m_g^2 \phi/p$ \cite{ehhs}.  
These terms are important when
$m_g^2 \phi/p \sim p^2$, or $p \sim m_g^{2/3} \phi^{1/3}$.
This is exactly the same scale at which Landau damping operates.
These effects will not alter the coefficient of the logarithmic
divergence (and hence the exponent), but they will produce terms
of order one in the gap equation, which contribute to $b_0'$ and
$b_1'$. 

{\bf (iv) Damping of the condensate:}\\  In the above, we neglected the
fact that the gap function has an imaginary part.  To
understand this imaginary part, consider first the self energy
for a quark in a Fermi sea.  As computed by
Le Bellac and Manuel, and by Vanderheyden and Ollitrault \cite{damping},
away from the edge of the Fermi sea, 
the quark can decay into another quark and a very soft gluon.
This is only possible with a HDL-resummed gluon, whose
spectral representation has support
from Landau damping in the space-like region.  The damping rate
of the quark behaves as $\sim g^2 |p - \mu|$, vanishing at the 
Fermi surface.  

{}From a similar physical process, the gap function acquires a
nonzero imaginary part when its momentum is away from the Fermi surface.
A quark can scatter into a quark with a different momentum through a
very soft gluon.  We can estimate the resulting
imaginary part of the gap function as follows \cite{prlett2}. 
If we had not restricted our analysis to
the principal value part of the energy denominators arising in
(\ref{matsu}), instead of $\ln |\epsilon_q^2 - \epsilon_k^2|$ in
Eq.\ (\ref{gapeqfinal}) we would have obtained 
$\ln (\epsilon_q^2-\epsilon_k^2)$. This logarithm has a cut
for $\epsilon_q < \epsilon_k$, giving rise to an imaginary part
for $\phi_k$,
\begin{equation}
{\rm Im}\, \phi_k \sim \bar{g}^2 \int_{\phi_0}^{\epsilon_k} 
\frac{d\epsilon_q}{\epsilon_q}\; \phi_q 
\simeq \bar{g}^2 \, \ln \left(\frac{\epsilon_k}{\phi_0}\right)\, \phi_0\,\, .
\end{equation}
Taking $\epsilon_k \sim b \mu \, e^{-x}$, cf.\ Eq.\ (\ref{defvarx}), 
momenta exponentially close to the Fermi surface occur when $x \sim x^*
= \pi/(2 \bar{g})$ (in weak coupling). 
In this region, the imaginary part of the gap function,
${\rm Im}\, \phi_k \sim  \bar{g}^2\, (x^*-x)\, \phi_0$, is {\it down\/}
by $\bar{g}$ relative to the real part,
${\rm Re}\, \phi_k \sim \sin(\bar{g}x)\, \phi_0$.
At the Fermi surface itself, $x = x^*$, the imaginary part
vanishes. Away from the Fermi surface, 
$\epsilon_k \sim \mu$, so $x \sim 1 $, and
$\phi_k$ is strongly damped, with the real and imaginary parts
of comparable magnitude, 
$ {\rm Re}\, \phi_k \sim {\rm Im}\, \phi_k \sim \bar{g}\, \phi_0$.

A gap function with a nonzero imaginary part is actually well
known from strongly coupled superconductors, as studied in
Eliashberg theory \cite{Scalapino}.  Damping occurs for
a similar reason as here, due to a nonzero imaginary part for
the plasmon.  

There is no problem in principle with including the damping of the gap
function.  A spectral representation for the gap function is
introduced, analogous to that of the quark and gluon propagators.
The resulting gap integrals are more involved (especially at
nonzero temperature), but can be treated in the manner which
we employed above.  

{\bf (v) Magnetic mass:} \\
As argued in Section I, at zero temperature the scale for the magnetic mass is
$\sim \mu \, \exp(-1/g^2)$. It is therefore
negligible compared to the scale of the condensate. At nonzero temperature,
the scale is no larger than $g^2 T$. For $T \sim \phi_0$, this is
down by $g^2$, and will only affect the prefactor of the gap to higher order
in $g$.

We conclude by stressing that the determination
of the prefactor is not merely an interesting problem in
its own right, but because it truly determines the physics
of color superconductivity (at least in weak coupling).
To the order at which we
compute, there is absolutely {\it no\/} preference for
the condensate to favor spin-zero over spin-one (or spin-two, {\it etc.}~!).
Surely the spin-one condensate is less favored than spin-zero;
the ratio of the two condensates is, in weak coupling, a pure
number which can be uniquely computed, once one knows how to
compute the prefactor in the condensate.

Indeed, perhaps one should entertain a more speculative hypothesis.
Even if a $J=0$ gap is favored, maybe there is always some
small admixture of higher-spin gaps,
and rotational invariance is inevitably broken in the true
ground state of color superconductivity.
\\ ~~ \\ ~~ \\
{\bf Acknowledgment}
\\ ~~ \\
We acknowledge discussions with D.\ Blaschke, W.\ Brown,
V.J.\ Emery, D.K.\ Hong, S.D.H.\ Hsu, M.\ Laine, J.T.\ Liu, 
V.N.\ Muthukumar, K.\ Rajagopal, H.C.\ Ren, T.\ Sch\"afer, and D.T.\ Son.
We especially thank T.\ Sch\"afer and D.T.\ Son for discussions 
on the ratio $T_c/\phi_0$.
R.D.P.\ was supported in part by DOE grant DE-AC02-98CH10886.
D.H.R.\ thanks RIKEN, BNL and the U.S.\ Dept.\ of Energy for
providing the facilities essential for the completion of this work,
and Columbia University's Nuclear Theory Group for
continuing access to their computing facilities.

\appendix

\section{The $J^P=1^+$ gaps} \label{A}

According to the results of subsection \ref{parity}, if
condensation occurs exclusively in the $J^P=1^+$ channel, the ansatz for 
the gap matrix reads (we suppress the dependence of the
gap functions on $K$ in the following)
\begin{equation}
J^P = 1^+\,:\;\;\; \Phi^+  =  \sum_{e=\pm}
\left[\bbox{\phi}_+^e \cdot \left( \hat{\bf k} + \bbox{\gamma}_\perp
({\bf k}) \right) \, \frac{1+e\,\gamma_0\, \bbox{\gamma} \cdot \hat{\bf k}}{2}
+
\bbox{\pi}_+^e \cdot \left( \hat{\bf k} + \bbox{\gamma}_\perp
({\bf k}) \right) \, \frac{1-e\,\gamma_0\, \bbox{\gamma} \cdot \hat{\bf k}}{2}
\right] \, \Lambda^e ({\bf k})\,\, .
\end{equation}
Now use
\begin{eqnarray}
\left(\hat{\bf k} + \bbox{\gamma}_\perp({\bf k}) \right) \,
\frac{1\pm e\, \gamma_0 \, \bbox{\gamma} \cdot \hat{\bf k}}{2}\, \Lambda^e
({\bf k}) & = & \hat{\bf k} \, \frac{1 \mp e\, \gamma_0}{4} 
\left[ (1 \pm \beta_{\bf k}\mp \alpha_{\bf k})
- (1 \pm \beta_{\bf k} \pm \alpha_{\bf k}) \, \bbox{\gamma} \cdot \hat{\bf k} 
\right] \nonumber \\
& + & \bbox{\gamma} \, \frac{1}{4}\, \left[(1 \pm \beta_{\bf k})\, (1 \pm e \,
\gamma_0\, \bbox{\gamma} \cdot \hat{\bf k}) \mp \alpha_{\bf k}\,
\bbox{\gamma} \cdot \hat{\bf k} + e\, \alpha_{\bf k} \, \gamma_0 \right] 
\end{eqnarray}
to obtain
\begin{eqnarray}
\Phi^+ & = & 
\left[ \frac{1+ \beta_{\bf k} - \alpha_{\bf k}}{4} 
\, (\bbox{\phi}_+^+ + \bbox{\phi}_+^-)
+      \frac{1- \beta_{\bf k} + \alpha_{\bf k}}{4} 
\, (\bbox{\pi}_+^+ + \bbox{\pi}_+^-) \right] \cdot \hat{\bf k} \nonumber \\
& + & 
\left[ - \frac{1+ \beta_{\bf k} - \alpha_{\bf k}}{4} 
\, (\bbox{\phi}_+^+ - \bbox{\phi}_+^-)
+      \frac{1- \beta_{\bf k} + \alpha_{\bf k}}{4} 
\, (\bbox{\pi}_+^+ - \bbox{\pi}_+^-) \right]\cdot \hat{\bf k} \, \gamma_0 
\nonumber \\
& + & 
\left[ - \frac{1+ \beta_{\bf k} + \alpha_{\bf k}}{4} 
\, (\bbox{\phi}_+^+ + \bbox{\phi}_+^-)
-      \frac{1- \beta_{\bf k} - \alpha_{\bf k}}{4} 
\, (\bbox{\pi}_+^+ + \bbox{\pi}_+^-) \right]\cdot \hat{\bf k} \, 
\bbox{\gamma} \cdot \hat{\bf k} \nonumber \\
& + & 
\left[  \frac{1+ \beta_{\bf k} + \alpha_{\bf k}}{4} 
\, (\bbox{\phi}_+^+ - \bbox{\phi}_+^-)
-      \frac{1- \beta_{\bf k} - \alpha_{\bf k}}{4} 
\, (\bbox{\pi}_+^+ - \bbox{\pi}_+^-) \right]\cdot \hat{\bf k} \, \gamma_0 \,
\bbox{\gamma} \cdot \hat{\bf k} \nonumber \\
& + & 
\left[ \frac{1+ \beta_{\bf k}}{4} 
\, (\bbox{\phi}_+^+ + \bbox{\phi}_+^-)
+      \frac{1- \beta_{\bf k} }{4} 
\, (\bbox{\pi}_+^+ + \bbox{\pi}_+^-) \right] \cdot \bbox{\gamma}  \nonumber \\
& + & \frac{\alpha_{\bf k}}{4} \,
\left( \bbox{\phi}_+^+ - \bbox{\phi}_+^-
+ \bbox{\pi}_+^+ - \bbox{\pi}_+^- \right)\cdot \bbox{\gamma}  \, \gamma_0 
\nonumber \\
& + &  \frac{\alpha_{\bf k}}{4} \,
\left( - \bbox{\phi}_+^+ - \bbox{\phi}_+^-
+ \bbox{\pi}_+^+ + \bbox{\pi}_+^- \right)\cdot \bbox{\gamma} \, 
\bbox{\gamma} \cdot \hat{\bf k} \nonumber \\
& + & 
\left[  \frac{1+ \beta_{\bf k}}{4} 
\, (\bbox{\phi}_+^+ - \bbox{\phi}_+^-)
-      \frac{1- \beta_{\bf k}}{4} 
\, (\bbox{\pi}_+^+ - \bbox{\pi}_+^-) \right]\cdot \bbox{\gamma} \, \gamma_0 \,
\bbox{\gamma} \cdot \hat{\bf k} \,\, . \label{explicitJ1}
\end{eqnarray}
Bailin and Love's ansatz for a $J^P=1^+$ gap reads [cf.\ Eq.\ (4.24) of
\cite{bailinlove}; we again suppress the momentum dependence of the gap
functions]
\begin{eqnarray}
\Delta & = & {\bf \Delta}_1 \cdot \bbox{\gamma} + {\bf \Delta}_2 
\cdot \hat{\bf k} \, \bbox{\gamma} \cdot \hat{\bf k}
+ i\, {\bf \Delta}_3 \times \hat{\bf k} \cdot \bbox{\gamma}\, \gamma_5
+ {\bf \Delta}_4 \cdot \hat{\bf k} \nonumber \\
& + & {\bf \Delta}_5 \cdot \bbox{\gamma} \, \gamma_0 + {\bf \Delta}_6 \cdot
\hat{\bf k} \, \bbox{\gamma} \cdot \hat{\bf k} \, \gamma_0
+ {\bf \Delta}_7 \cdot \hat{\bf k}\, \gamma_0 + i\,
{\bf \Delta}_8 \times \hat{\bf k} \cdot \bbox{\gamma} \, \gamma_0\, 
\gamma_5 \,\, . \label{bl}
\end{eqnarray}
[We added an $i$ in the last term as compared to \cite{bailinlove}.
This simplifies the notation in the following, but is not essential, 
as the gap functions are in general complex-valued.]
With the definition of $\gamma_5 = i \gamma_0 \gamma^1 \gamma^2
\gamma^3$ one computes
\begin{equation} \label{spatprodukt}
i\, {\bf \Delta} \times \hat{\bf k} \cdot \bbox{\gamma}\, \gamma_5
= - \gamma_0 \left( {\bf \Delta} \cdot \hat{\bf k} +
{\bf \Delta} \cdot \bbox{\gamma} \, \bbox{\gamma} \cdot \hat{\bf k}
\right) \equiv - \gamma_0\, {\bf \Delta} \cdot \bbox{\gamma}_\perp
({\bf k}) \, \bbox{\gamma} \cdot \hat{\bf k}\,\,,
\end{equation}
and rewrites (\ref{bl}) as
\begin{eqnarray}
\Delta & = & ({\bf \Delta}_4 + {\bf \Delta}_8) \cdot \hat{\bf k}
+ ({\bf \Delta}_7 - {\bf \Delta}_3) \cdot \hat{\bf k}\, \gamma_0
+ {\bf \Delta}_2 \cdot \hat{\bf k} \, \bbox{\gamma} \cdot \hat{\bf k}
- {\bf \Delta}_6 \cdot \hat{\bf k} \, \gamma_0 \,
                 \bbox{\gamma} \cdot \hat{\bf k} \nonumber \\
& + & {\bf \Delta}_1 \cdot \bbox{\gamma}
+ {\bf \Delta}_5 \cdot \bbox{\gamma} \, \gamma_0 
+ {\bf \Delta}_8 \cdot \bbox{\gamma} \, \bbox{\gamma} \cdot \hat{\bf k}
+ {\bf \Delta}_3 \cdot \bbox{\gamma}\, \gamma_0 \, \bbox{\gamma}
\cdot \hat{\bf k}\,\, . \label{bl2}
\end{eqnarray}
Direct comparison of (\ref{explicitJ1}) and (\ref{bl2}) reveals
\begin{mathletters} \label{relations}
\begin{eqnarray}
{\bf \Delta}_1 & = & {\bf \Delta}_4 = \frac{1+ \beta_{\bf k}}{4} 
\, (\bbox{\phi}_+^+ + \bbox{\phi}_+^-)
+      \frac{1- \beta_{\bf k} }{4} 
\, (\bbox{\pi}_+^+ + \bbox{\pi}_+^-) \,\, , \\
{\bf \Delta}_2 & = &  - {\bf \Delta}_1 + {\bf \Delta}_8 \,\, , \\
{\bf \Delta}_3 & = &  \frac{1+ \beta_{\bf k}}{4} 
\, (\bbox{\phi}_+^+ - \bbox{\phi}_+^-)
-      \frac{1- \beta_{\bf k}}{4} 
\, (\bbox{\pi}_+^+ - \bbox{\pi}_+^-) \,\, , \\
{\bf \Delta}_5 & = & {\bf \Delta}_7 = \frac{\alpha_{\bf k}}{4} \,
\left( \bbox{\phi}_+^+ - \bbox{\phi}_+^-
+ \bbox{\pi}_+^+ - \bbox{\pi}_+^- \right) \,\, , \\
{\bf \Delta}_6 & = & - {\bf \Delta}_3 - {\bf \Delta}_5 \,\, , \\
{\bf \Delta}_8 & = & \frac{\alpha_{\bf k}}{4} \,
\left( - \bbox{\phi}_+^+ - \bbox{\phi}_+^-
+ \bbox{\pi}_+^+ + \bbox{\pi}_+^- \right)\,\, .
\end{eqnarray}
\end{mathletters}
These relations exhibit a redundancy in Bailin and Love's 
ansatz (\ref{bl}): only four of the eight 3-vector gap
functions ${\bf \Delta}_1, \ldots, {\bf \Delta}_8$ are
independent. The reason for this redundancy is
that in Eq.\ (\ref{bl}) the transverse components of
${\bf \Delta}_2,\, {\bf \Delta}_4,\, {\bf \Delta}_6,$ and
${\bf \Delta}_7$ never appear, as these gap functions are 
projected in the longitudinal direction $\hat{\bf k}$.
Furthermore, only the transverse components of
${\bf \Delta}_3$ and ${\bf \Delta}_8$ appear
on account of (\ref{spatprodukt}).
Finally, the longitudinal components of the 
remaining two gap functions ${\bf \Delta}_1$ and
${\bf \Delta}_5$ can be absorbed by redefining ${\bf \Delta}_2$ and
${\bf \Delta}_6$, thus only their transverse components
contribute. In this manner, only half of the original 24 gap functions
are independent. 
Physically, this can be understood from the restriction to
the positive-parity channel $J^P=1^+$.
Thus, only the four 3-vectors $\bbox{\phi}_+^\pm, \, \bbox{\pi}_+^\pm$ 
appear on the right-hand side of (\ref{relations}). The other
four 3-vectors $\bbox{\phi}_-^\pm, \, \bbox{\pi}_-^\pm$ do
not contribute, as they correspond to pairing in the
$J^P= 1^-$ channel. 

This said, one can readily derive
a more efficient form of Bailin and Love's ansatz (\ref{bl}), 
which utilizes both longitudinal and transverse components
of the truly independent gap functions. Choosing the
latter to be ${\bf \Delta}_1,\, {\bf \Delta}_3,\, {\bf \Delta}_5$,
and ${\bf \Delta}_8$, we obtain from (\ref{bl}) with
(\ref{relations})
\begin{equation}
\Delta  =  {\bf \Delta}_1 \cdot \left(\hat{\bf k}
+ \bbox{\gamma}_\perp({\bf k}) \right) + {\bf \Delta}_3
\cdot \left(\hat{\bf k} + \bbox{\gamma}_\perp({\bf k}) \right)
\, \gamma_0 \, \bbox{\gamma} \cdot \hat{\bf k}
+ {\bf \Delta}_5 \cdot \left(\hat{\bf k} + 
\bbox{\gamma}_\perp({\bf k}) \right) \gamma_0
+ {\bf \Delta}_8 \cdot 
\left(\hat{\bf k} + \bbox{\gamma}_\perp({\bf k}) \right) 
\, \bbox{\gamma} \cdot \hat{\bf k}\,\, . \label{bl3}
\end{equation}
It is now also easy to interpret the results obtained by
Bailin and Love in \cite{bailinlove2}, where they studied
$J^P = 1^+$ condensation in the ultrarelativistic limit.
In this limit, $\alpha_{\bf k} = 0$, thus
${\bf \Delta}_5 = {\bf \Delta}_8=0$ on account
of (\ref{relations}). Bailin and Love discuss two order parameters
for condensation in the $J^P=1^+$ channel, the first
being the longitudinal component of
${\bf \Delta}_4 - {\bf \Delta}_5 - {\bf \Delta}_6$, the second 
being $({\bf \Delta}_1 + {\bf \Delta}_3) \cdot
(\bbox{\gamma} - \hat{\bf k}\, \bbox{\gamma} \cdot \hat{\bf k}
- i\, \hat{\bf k} \times \bbox{\gamma}\, \gamma_5)$.

{}From (\ref{relations}), we identify the first with
the longitudinal component of ${\bf \Delta}_1 + {\bf \Delta}_3$. From 
(\ref{spatprodukt}), we realize that the second 
is identical to $({\bf \Delta}_1 + {\bf \Delta}_3) \cdot
\bbox{\gamma}_\perp ({\bf k})\, 
(1-\gamma_0\, \bbox{\gamma} \cdot \hat{\bf k})$, {\it i.e.},
essentially the transverse components of ${\bf \Delta}_1 + {\bf \Delta}_3$.
Thus, Bailin and Love discuss two separate gap equations, one
for the longitudinal, the other for the transverse components of the
independent gap function ${\bf \Delta}_1 + {\bf \Delta}_3$.
Although we also find two gap equations for the longitudinal and
the transverse gaps, cf.\ Section \ref{J1}, we do {\em not\/} find
that they decouple, as the excitation energies (\ref{quasipenergy})
contain {\em both\/} longitudinal as well as transverse gap functions.


\begin{thebibliography}{99}

\bibitem{BCS} 
J.R.\ Schrieffer, {\it Theory of Superconductivity}
(New York, W.A.\ Benjamin, 1964).

\bibitem{fetter}
A.L.\ Fetter and J.D.\ Walecka, {\it Quantum Theory of Many-Particle Systems}
(McGraw--Hill, New York, 1971); A.A.\ Abrikosov, L.P.\ Gorkov, and I.E.\ 
Dzyaloshinski, {\it Methods of Quantum Field Theory in Statistical 
Physics} (Dover, New York, 1963).

\bibitem{Scalapino}
D.J.\ Scalapino, in: {\it Superconductivity}, ed.\ R.D.\ Parks,
(New York, M. Dekker, 1969), p.\ 449ff.

\bibitem{bailinlove}
D.\ Bailin and A.\ Love, Phys.\ Rep.\ {\bf 107}, 325 (1984).

\bibitem{general}
M.\ Alford, K.\ Rajagopal, and F.\ Wilczek, Phys.\ Lett.\ {\bf B422},
247 (1998); 
R.\ Rapp, T.\ Sch\"afer, E.V.\ Shuryak, and M.\ Velkovsky,
Phys.\ Rev.\ Lett.\ {\bf 81}, 53 (1998); hep-ph/9904353;
N.\ Evans, S.D.H.\ Hsu, and M.\ Schwetz, Nucl.\ Phys.\ {\bf B551}, 
275 (1999); Phys. Lett. {\bf B449} 281, (1999);
J.\ Berges and K.\ Rajagopal, Nucl.\ Phys.\ {\bf B538}, 215 (1999);
T.\ Sch\"afer and F.\ Wilczek, Phys.\ Lett.\ {\bf B450}, 325 (1999); 
G.W.\ Carter and D.\ Diakonov, Phys.\ Rev.\ D {\bf 60}, 016004 (1999);
K.\ Langfeld and M.\ Rho, hep-ph/9811227;
M.\ Alford, J.\ Berges, and K.\ Rajagopal, hep-ph/9903502;
E.\ Shuster and  D.T.\ Son, hep-ph/9905448;
D.K.\ Hong, M.\ Rho, and I.\ Zahed, hep-ph/9906551;
V.A.\ Miransky, I.A.\ Shovkovy, and L.C.R.\ Wijewardhana, hep-ph/9908212;
R.\ Casalbuoni and R.\ Gatto, hep-ph/9908227; hep-ph/9909419; 
B.Y.\ Park, M.\ Rho, A.\ Wirzbad, and I.\ Zahed, SUNYSB-preprint
(unpublished).

\bibitem{colorflavorlock}
M.\ Alford, K.\ Rajagopal, and F.\ Wilczek, Nucl.\ Phys.\
{\bf B537}, 443 (1999).

\bibitem{prlett}
R.D.\ Pisarski and D.H.\ Rischke, Phys.\ Rev.\ Lett.\ {\bf 83}, 37 (1999).

\bibitem{Son}
D.T.\ Son, Phys.\ Rev.\ D {\bf 59}, 094019 (1999).

\bibitem{prscalar}
R.D.\ Pisarski and D.H.\ Rischke, Phys.\ Rev.\ D {\bf 60}, 094013 (1999).

\bibitem{hong}
D.K.\ Hong, hep-ph/9812510, hep-ph/9905523.

\bibitem{continuity}
T.\ Sch\"afer and F.\ Wilczek, Phys.\ Rev.\ Lett.\ {\bf 82}, 3956 (1999);
hep-ph/9903503.

\bibitem{hongetal}
D.K.\ Hong, V.A.\ Miransky, I.A.\ Shovkovy, and L.C.R.\ Wijewardhana,
hep-ph/9906478.

\bibitem{SchaferWilczek}
T.\ Sch\"afer and F.\ Wilczek, hep-ph/9906512.

\bibitem{parity}
R.D.\ Pisarski and D.H.\ Rischke, nucl-th/9906050.

\bibitem{prlett2}
R.D.\ Pisarski and D.H.\ Rischke, nucl-th/9907041.

\bibitem{JMEproc}
R.D.\ Pisarski and D.H.\ Rischke, nucl-th/9907094.

\bibitem{rockefeller}
W.E.\ Brown, J.T.\ Liu, and H.-C.\ Ren, hep-ph/9908248.

\bibitem{hsuschwetz}
S.D.H.\ Hsu and M.\ Schwetz, hep-ph/9908310.

\bibitem{schaferlast}
T.\ Sch\"afer, hep-ph/9909574; I.A.\ Shovkovy and
L.C.R.\ Wijewardhana, hep-ph/9910225.

\bibitem{ehhs}
N.\ Evans, J.\ Hormuzdiar, S.D.H.\ Hsu, and M.\ Schwetz,
hep-ph/9910313.

\bibitem{temp}
J.O.\ Andersen, E.\ Braaten, and M.\ Strickland, hep-ph/9902327,
hep-ph/9905337;
J.-P.\ Blaizot, E.\ Iancu, and A.\ Rebhan, hep-ph/9906340.

\bibitem{freenergy}
B.\ Freedman and L.D.\ McLerran, Phys.\ Rev.\ D 
{\bf 16}, 1130 (1977); {\bf 16}, 1147 (1977); {\bf 16}, 1169 (1977);
{\bf 17}, 1109 (1978); R.\ Baier and K.\ Redlich, hep-ph/9908372.

\bibitem{RDPphysica}
R.D.\ Pisarski, Physica {\bf A 158}, 146 (1989).

\bibitem{LeBellac}
M.\ Le Bellac, {\it Thermal Field Theory}
(Cambridge, Cambridge University Press, 1996).

\bibitem{finitedens}
J.-P.\ Blaizot and J.-Y.\ Ollitrault, Phys.\ Rev.\ D {\bf 48}, 1390 (1993);
H.\ Vija and M.H.\ Thoma, Phys.\ Lett.\ {\bf B342}, 212 (1995);
C.\ Manuel, Phys.\ Rev.\ D {\bf 53}, 5866 (1996).

\bibitem{damping}
M.\ Le Bellac and C.\ Manuel, Phys.\ Rev.\ D {\bf 55}, 3215 (1997);
B.\ Vanderheyden and J.-Y.\ Ollitrault, Phys.\ Rev.\ D {\bf 56}, 5108 (1997).

\bibitem{kkr}
R.\ Kobes, G.\ Kunstatter, and A.\ Rebhan, 
Phys.\ Rev.\ Lett.\ {\bf 64}, 2992 (1990);
Nucl.\ Phys.\ {\bf B355}, 1 (1991).

\bibitem{holstein}
T.\ Holstein, R.E.\ Norton, and P.\ Pincus, Phys.\ Rev.\ B {\bf 6},
2649 (1973).

\bibitem{bailinlove2} 
D.\ Bailin and A.\ Love, Nucl.\ Phys.\ {\bf B190} [FS3], 175 (1981).
\end{thebibliography}
\end{document}